\documentclass[]{ceurart}
\usepackage{listings}
\usepackage{longtable}
\usepackage{lscape}
\usepackage{geometry}
\usepackage{float}
\usepackage[justification=centering]{caption}
\usepackage[T1]{fontenc} 
\usepackage[utf8]{inputenc} 
\usepackage{ifthen}
\usepackage{graphicx}
\usepackage{hyperref}
\usepackage{multirow}
\usepackage{array} 
\usepackage{arydshln} 
\usepackage[many]{tcolorbox} 

\usepackage{xspace}
\usepackage{xcolor}
\usepackage[normalem]{ulem} 			
\usepackage{amsmath,amsfonts}

\usepackage{caption} 
\usepackage{subcaption} 

\newcolumntype{L}[1]{>{\raggedright\let\newline\\\arraybackslash\hspace{0pt}}p{#1}}
\newcolumntype{C}[1]{>{\centering\let\newline\\\arraybackslash\hspace{0pt}}p{#1}}
\newcolumntype{R}[1]{>{\raggedleft\let\newline\\\arraybackslash\hspace{0pt}}p{#1}}

\newboolean{showcomments}
\setboolean{showcomments}{true}

\ifthenelse{\boolean{showcomments}}
{
	\newcommand{\nb}[3]{
		{\colorbox{#2}{\bfseries\sffamily\scriptsize\textcolor{white}{#1}}}
		{\textcolor{#2}{\sf\small$\blacktriangleright$\textit{#3}$\blacktriangleleft$}}}
	 
	\newcommand{\bnote}[2]{\fbox{\color{blue}\bfseries\sffamily\scriptsize#1}
    	{\color{blue}\sf\small$\blacktriangleright$\textit{#2}$\blacktriangleleft$}}
	\newcommand{\old}[1]{{\color{gray}\sout{#1}}} 
	\newcommand{\del}[1]{\old{#1}} 
	\newcommand{\ins}[1]{{\textcolor{blue}{\uline{#1}}}} 
	\newcommand{\ugh}[1]{{\textcolor{red}{\uwave{#1}}}} 
	\newcommand{\chg}[2]{{\textcolor{red}{\sout{#1}}}{\ra}\textcolor{blue}{\uline{#2}}} 
	 
	\newcommand{\fix}[1]{\bnote{FIX}{#1}}
}{
	\newcommand{\bnote}[2]{}
	\newcommand{\nb}[3]{}
	\newcommand{\old}[1]{}
	\newcommand{\del}[1]{}
	\newcommand{\ins}[1]{}
	\newcommand{\ugh}[1]{}
	\newcommand{\chg}[2]{}
	
	\newcommand{\fix}[1]{}
} 

\newcommand{\hide}[1]{}

\usepackage[many]{tcolorbox}  

\newcommand{\comp}{Complishon }

\graphicspath{{figures/}{newFigures/}}

\newcommand{\commented}[1]{}

\newcommand{\eg}{\emph{e.g.,}\xspace}

\newcommand{\etal}{\emph{et al.,}\xspace}
\newcommand{\ct}[1]{{\textsf{#1}}\xspace}

\usepackage{url}            
\makeatletter
\def\url@leostyle{%
  \@ifundefined{selectfont}{\def\UrlFont{\sf}}{\def\UrlFont{\small\sffamily}}}
\makeatother
\definecolor{main}{HTML}{828282}    
\definecolor{sub}{HTML}{E0E0E0}     

\tcbset{
    sharp corners,
    colback = white,
    before skip = 0.2cm,    
    after skip = 0.5cm      
}                           

\newtcolorbox{cbox}{
    enhanced, 
    boxrule = 0pt, 
    borderline = {0.75pt}{0pt}{main}, 
    borderline = {0.75pt}{2pt}{sub} 
}

\newcommand{\compl}{\textsc{Complishon}\xspace}

\begin{document}
\copyrightyear{2025}
\copyrightclause{Copyright for this paper by its authors. Use permitted under Creative Commons License Attribution 4.0 International (CC BY 4.0).}

\conference{IWST 2025: International Workshop on Smalltalk Technologies, July 1-4,  2025, Gdańsk, Poland}

\title{Package-Aware Approach for Repository-Level Code Completion in Pharo}

\address[1]{Univ. Lille, Inria, CNRS, Centrale Lille, UMR 9189 CRIStAL, Park Plaza, Parc scientifique de la Haute-Borne, 40 Av. Halley Bât A, 59650 Villeneuve-d’Ascq, France}

\address[2]{CIRAD, UMR SENS, F-34000 Montpellier, France}

\address[3]{CNRS, University of Bordeaux, Bordeaux INP, LaBRI, UMR5800, F-33400, Talence, France}

\author[1]{Omar AbedelKader}[%
email=omar.abedelkader@inria.fr,%
orcid=0009-0005-1339-5683,%
url=https://omarabedelkader.github.io%
]

\author[1]{Stéphane Ducasse}[%
email=stephane.ducasse@inria.fr,%
orcid=0000-0001-6070-6599,%
url=https://stephane.ducasse.free.fr/%
]

\author[2]{Oleksandr Zaitsev}[%
email=oleksandr.zaitsev@cirad.fr,%
orcid=0000-0003-0267-2874,%
url=https://umr-sens.fr/fr/-/zaitsev-oleksandr%
]

\author[3]{Romain Robbes}[%
email=romain.robbes@labri.fr,%
orcid=0000-0003-4569-6868,%
url=https://rrobbes.github.io/,%
]

\author[1]{Guillermo Polito}[%
email=guillermo.polito@inria.fr,%
orcid=0000-0003-0813-8584,%
url=https://guillep.github.io,%
]

\begin{abstract}
Pharo offers a sophisticated completion engine based on semantic heuristics, which coordinates specific fetchers within a lazy architecture. These heuristics can be recomposed to support various activities (\eg live programming or history usage navigation). While this system is powerful, it does not account for the repository structure when suggesting global names such as class names, class variables, or global variables. As a result, it does not prioritize classes within the same package or project, treating all global names equally.

In this paper, we present a new heuristic that addresses this limitation. Our approach searches variable names in a structured manner: it begins with the package of the requesting class, then expands to other packages within the same repository, and finally considers the global namespace. We describe the logic behind this heuristic and evaluate it against the default semantic heuristic and one that directly queries the global namespace. Preliminary results indicate that the Mean Reciprocal Rank (MRR) improves, confirming that package-awareness completions deliver more accurate and relevant suggestions than the previous flat global approach.
\end{abstract}


\maketitle

\section{Introduction}\label{introduction}

Modern code completion systems are increasingly expected to be intelligent, predicting not only syntactically valid tokens but also semantically relevant ones, thereby reducing the effort required to navigate large codebases \cite{Jin25a}. In the Pharo programming environment \cite{Blac09a}, \comp is a sophisticated code completion engine designed with these goals. Developed by the last author, \comp employs a modular and lazy architecture based on semantic heuristics that consider both the type of entity (\eg variable, class, method) and its location in the program structure (\eg prioritizing classes in the current class hierarchy before superclasses). Unlike traditional alphabetical or purely syntactic completion systems, such as those criticized by Bruch \etal \cite{Bruc09a} and Robbes \etal \cite{Robb10c}, \compl is grounded in the idea that completion quality can be significantly improved by taking into account a context.

However, even with these heuristics, \compl, like many completion engines, originally treated the global namespace as flat, walking through global entities in a non-deterministic order other than matching prefixes. This becomes problematic in industrial-scale systems, where the number of possible completions can be overwhelming. For instance, companies using Pharo in production, such as Lifeware, maintain systems exceeding 30 million lines of code \footnote{\href{https://www.inria.fr/en/pharo-programming-language-celebrates-its-10th-anniversary-after-going-around-world}{10th Anniversary of Pharo}}. In these contexts, completion suggestions must not only be semantically relevant but also efficiently scoped to avoid information overload. One significant omission in the original design of \compl was awareness of package structure.

\begin{figure}[htbp]
\begin{center}
   \includegraphics[width=0.9\linewidth]{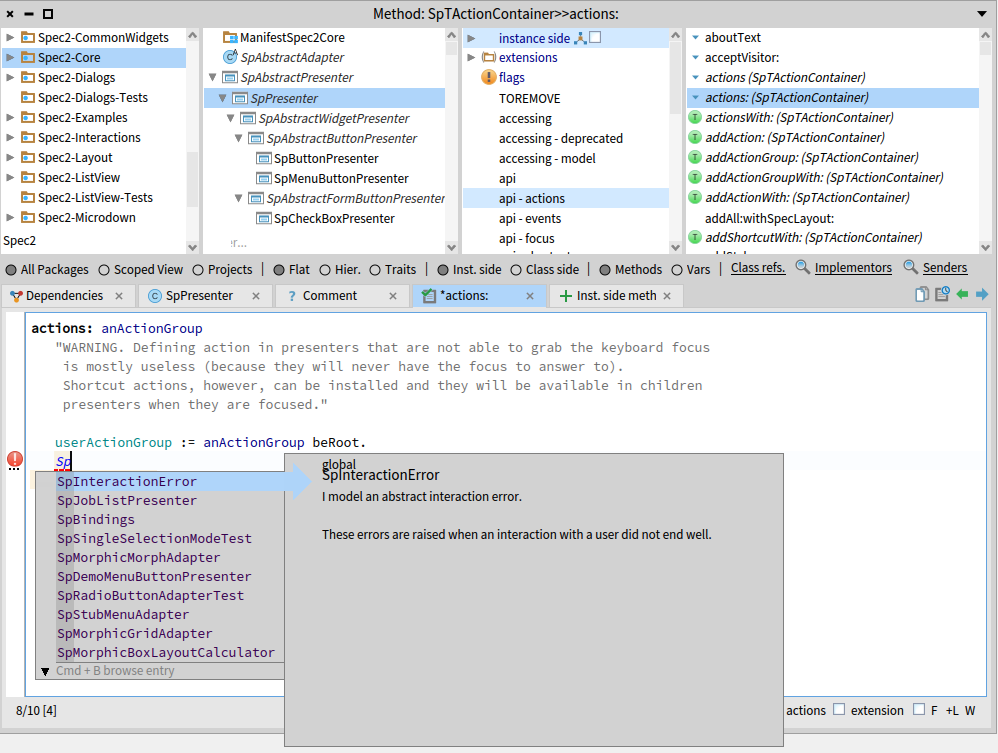}
\caption{\textbf{Default code completion interface in Pharo (Flat Global Scope Suggestions Without Package Awareness).}\label{fig:exflat}}
\end{center}
\end{figure}

While it excelled in suggesting global names like classes or variables, it did not favor entities from the same package or related parts of the system. 
\compl does not prioritize or even adequately surface entities from the same package as the current editing context. As shown in Figure \ref{fig:exflat}, when invoking code completion within the \ct{SpPresenter} class (located in the \ct{Spec2-Core} package), the suggestions are entirely drawn from the global namespace, ignoring nearby classes such as \ct{SpPresenterBuilder}, \ct{SpTextPresenter}, or \ct{SpApplication}, which are defined in the same package. This leads to a degraded developer experience, especially in large codebases where numerous globally accessible entities can easily overshadow locally relevant ones. 

Figure \ref{fig:exflat} shows the potential candidate completions from packages \ct{Spec2-core}, highlighting \compl's failure to leverage local package context effectively. This is critical in Pharo, where projects are modularized into packages (\eg \ct{Spec2-Core}, \ct{Spec2-Dialogs}, \ct{Spec2-Interaction}) that group functionally related classes. Developers often work within a small subset of the system, typically their packages, its dependencies, and core libraries, and should not be distracted by completions from unrelated parts of the codebase.  Specifically, the figure illustrates an error where the first ten completion suggestions (such as \ct{SpInteractionError}, \ct{SpJobListPresenter}, etc.) are not part of the \ct{Spec2-Core} package, thus underscoring the importance of package-aware code completion that respects package boundaries.

This article proposes and evaluates a repository-level completion strategy, a simple yet effective enhancement to the \compl architecture that ranks candidates from the current package highest, followed by suggestions from lateral packages, and only then from the rest global namespace. The goal is to improve both relevance and responsiveness by exploiting the modularity already present in Pharo subsystems. This approach resonates with trends in completion research that use structural or probabilistic models such as Bayesian strategies \cite{Prok15a} or JetBrains' log-based rankings \cite{Biba22a} to make completions more context-aware. Our work is further motivated by principles from the moldable development paradigm, as described by Chis \etal \cite{Chis15b}, which argues that tools like code completion should be extensible and adaptable to specific development contexts. In this sense, \compl aligns with these ideals by offering a plugin architecture for heuristics, including our proposed package-awareness logic.

This article is structured as follows: Section~\ref{background} gives an overview of the \compl engine and its modular heuristic-based architecture. Section~\ref{approach} identifies the limitations of global-environment-based completion, presents our approach, and outlines our hypothesis for package awareness suggestions. Section~\ref{evaluation} describes our evaluation across multiple projects and strategies. Section~\ref{discussion} discusses the findings and their implications for completion systems in dynamic, large-scale environments. Section~\ref{relatedwork} situates our work within the broader landscape of static code completion. Section~\ref{limitations} outlines the main limitations of our approach, including the reliance on static reference points, the use of truncated identifiers that may not reflect real-world usage, and the challenges posed by Pharo-specific naming and package structures. Finally, Section~\ref{conclusion} outlines future directions for integrating adaptive and learned strategies into the \compl engine.

\section{Background}\label{background}

\begin{figure}[htbp]
\begin{center}
   \includegraphics[width=0.9\linewidth]{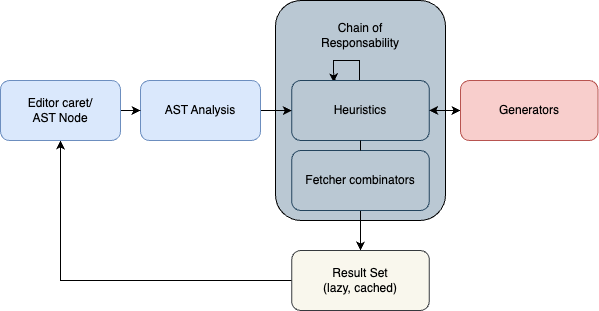}
\end{center}
\caption{\textbf{\compl's architecture}\label{fig:complishon}}
\end{figure}

 \compl the Pharo completion engine (see Figure~\ref{fig:complishon}), consists primarily of three key components: Heuristics, Lazy Fetchers, and a lazily cached Result Set. At the core, heuristics provide semantic guidance for the completion process by analyzing the Abstract Syntax Tree (AST) node located at the cursor (editor caret) and selecting the appropriate fetchers for completion suggestions. These heuristics are structured in a chain of responsibility \cite{Gamm95a}, allowing a heuristic to pass handling responsibilities down the chain if it cannot process the current AST node itself. Fetchers, implemented using combinators, lazily retrieve and filter potential code completion candidates based on the context and user input, significantly optimizing performance and memory usage. A decorator pattern further enhances fetchers by preventing duplicate suggestions, particularly crucial in scenarios involving method inheritance ensuring results remain relevant and unique. Fetchers utilize specialized filters (\eg \ct{CoBeginsWithFilter}), to match completion suggestions with the user's partially typed input. 
 
The Result Set component serves as a lazy, cached store that accumulates the suggestions provided by fetchers only as required, further enhancing efficiency. \compl's leverages AST-based analysis and utilizes a double-dispatch mechanism to adapt dynamically to the surrounding context. It constructs a dedicated context for code completion, considering factors such as source text and caret position.  The generation and presentation of suggestions are managed by the IDE, which also integrates strategies such as case sensitivity filtering and adaptive configuration based on the structure and semantics of the current code environment. This adaptability employs heuristic-based fetchers configured by visitor patterns, informed by parsing and typing processes, to refine output and dynamically eliminate redundant or irrelevant suggestions. 

The heuristics are modular, specialized for different code elements such as messages, variables, and symbols, and systematically connected in sequences forming a robust and comprehensive filtering framework. This chaining process includes sophisticated program semantics-based strategies such as prioritizing instance variables before superclass variables, self message suggestions, inherited methods, and inferred initialization constructs. Consequently, \compl provides an efficient, contextually aware, and highly accurate completion experience tailored precisely to the user's current coding scenario.

\section{Our approach: Repository level package structure}\label{approach}
\subsection{Repository Level Completion} \label{repolc}

Although \compl is effective in identifying global entities such as class names, class variables, or global variables, it currently does not leverage the package structure of a project effectively. As a result, it treats all global names uniformly, offering no preference to entities located in the same or related packages. This behavior contrasts with several insights from prior research, which emphasize the importance of contextual filtering and locality in improving the relevance of code completion results. For instance, Hou \etal \cite{Hou10a} propose heuristic techniques to filter and sort API suggestions using type hierarchies and grouping, significantly reducing the visual clutter of irrelevant entries. Similarly, Bruch \etal \cite{Bruc09a} introduced the Best Matching Neighbors (BMN) approach, which ranks suggestions based on similarity to local usage contexts, an idea analogous to prioritizing completions from the same or nearby packages. Robbes \etal \cite{Robb08a} further underscore the value of recent usage and lexical proximity, showing that local context often outperforms global frequency. More recently, Hellendoorn \etal \cite{Hell19a} demonstrated that intra-project completions remain the hardest for current models, largely due to their inability to distinguish between local and global identifiers. In a complementary direction, Li \etal \cite{Li21b} proposed learned acceptance and ranking models to suppress noisy completions, optimizing not just for correctness but for developer effort.

\begin{figure}[htbp]
\begin{center}
   \includegraphics[width=0.9\linewidth]{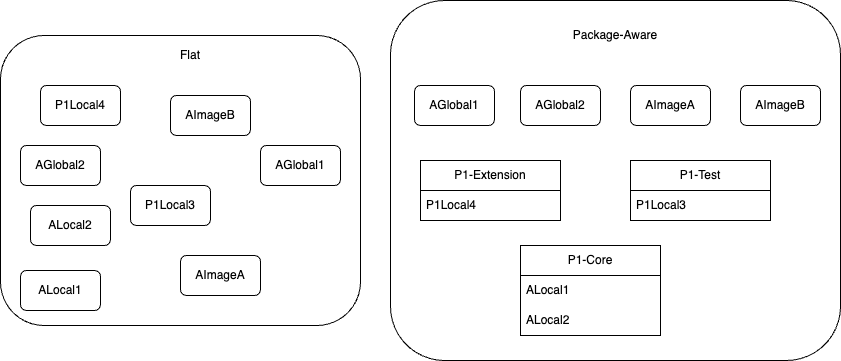}
\caption{\textbf{Flat vs.  Package Awarness}. In the flat model, all global variables are visible without a package context. In the package-aware model, name resolution follows a structured order: current package (P1-Core), then related packages (P1-Test, P1-Extension), and finally global namespace (AGlobal, AImage).\label{fig:flatvspackag}}
\end{center}
\end{figure}

Inspired by these works, we introduce a new heuristic in \compl that leverages package structure to improve completion prioritization (See Figure~\ref{fig:flatvspackag}). For example, when performing auto-completion within a class in the \ct{P1-Core} package, typing the letter A should ideally yield suggestions in the following order: first from \ct{P1-Core}, then from related packages such as \ct{P1-Extension} or \ct{P1-Test}, and only afterward from the global namespace.

\subsection{Leveraging Project Package Structure}

To enable this behavior, we designed a package-aware completion heuristic that operates in three steps:

\begin{itemize}

\item \textbf{Identifying the current package}: The completion engine determines the active package context, such as \ct{P1-Core}.
\item \textbf{Collecting potential matches}: Typing A triggers a scan for all relevant matches in local and global names.
\item \textbf{Prioritizing based on package proximity}:

\begin{enumerate}
	\item First, suggest entities defined within the current package (e.g., \ct{P1-Core}).
	\item Next, prefer suggestions from closely related packages (Lateral Dependencies) (e.g., \ct{P1-Extension} or \ct{P1-Test}). These are currently inferred from naming similarity rather than formal dependencies.
	\item Finally, include entities from the remaining global scope, if further suggestions are needed.
\end{enumerate}
\end{itemize}

This strategy mirrors user expectations by elevating locally relevant completions, thereby reducing cognitive load and supporting the findings of prior empirical and usability studies.

\subsection{Implementation Details}

\begin{figure}[htbp]
\begin{center}
   \includegraphics[width=0.9\linewidth]{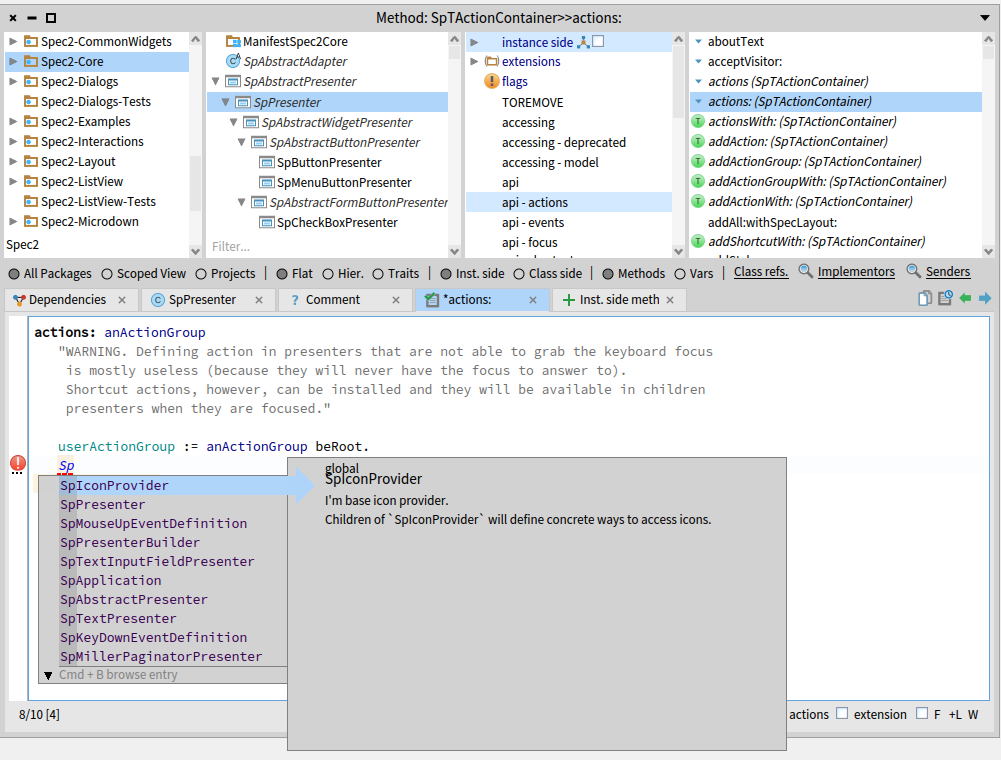}
\caption{\textbf{Improved Suggestions with Package-Scoped Heuristic}\label{fig:expackageScopedSuggestions}}
\end{center}
\end{figure}

This behavior is implemented through new extension points in \compl, enabling fine-grained control over completion fetchers and filtering logic. These enhancements support dynamic reordering of suggestions based on a package-locality heuristic, which prioritizes elements from the current package and its immediate relations. For instance, when auto-completing in a class within the package P1-Core, typing the letter A will first yield suggestions from P1-Core, followed by related packages like P1-Extension or P1-Test, and finally the global namespace. This structured prioritization reduces cognitive overhead and improves developer productivity. To evaluate the impact of this heuristic, we refined our benchmarks to focus on variable completions specifically. As shown in Figure \ref{fig:expackageScopedSuggestions}, invoking completion in a Spec2-Core context now highlights relevant classes such as SpIconProvider, SpPresenterBuilder, and SpTextPresenter. These results confirm the heuristic’s effectiveness in promoting package-local relevance and improving suggestion accuracy.

\section{Evaluation}\label{evaluation}
\subsection{Benchmark Logic}

In this work, we only focus on the completion of global variables such as class names \footnote{In Pharo, class names are global variables. In practice, most globals considered in this study are class names.}. To evaluate the effectiveness of variable name completion algorithms, we based our implementation on the benchmark methodology introduced by Robbes \etal \cite{Robb08a, Robb08b}. Although their original benchmark relied on a change-based repository of program history, our approach adapts the core idea for static analysis and variable-centric benchmarking, making it applicable in contexts where historical data is unavailable. The essence of this benchmark is to test whether a completion engine can correctly suggest the original names of variables when only partial prefixes are provided. The fundamental insight is that by systematically rewriting every variable access site with increasingly longer prefixes (ranging from 2 to 8 characters), we can invoke the completion engine and evaluate whether it ranks the correct name high in its suggestions. This method simulates realistic completion scenarios and allows us to measure not just correctness but also ranking performance and user effort reduction. Our system, implemented via the \ct{StaticBenchmarksVariables} class, carries out the following steps for each method in a given package:

\begin{enumerate}
\item \textbf{AST Extraction}: The method’s AST is retrieved to analyze variable usages within the code.
\item \textbf{Variable Filtering}: Each variable node is examined to confirm it is global (usually indicated by an uppercase-starting name).
\item \textbf{Name Masking}: The variable name is programmatically shortened to generate several prefixes of increasing lengths, from 2 up to 8 characters (or the full name length, if shorter). For example, the variable \ct{OrderedCollection} yields prefixes such as \ct{Or}, \ct{Ord}, \ct{Orde}, etc.
\item \textbf{Completion Invocation}: For each prefix, the completion engine is triggered as if the user were requesting suggestions after typing that fragment.
\item \textbf{Result Logging}: The engine’s output is analyzed to determine if the original name appears among the top 10 suggestions. If found, we record the rank at which it was suggested. If not, it is considered a failure for that prefix length.
\end{enumerate}

This process is repeated for all global variables in all methods of the targeted package or class scope. The benchmark is implemented in \ct{StaticBenchmarks} class and uses a \ct{ResultSetBuilder} to generate completions based on specified heuristics. We provide a modular API allowing execution across individual classes or entire packages. Several metrics are collected:

\begin{enumerate}

\item \textbf{Accuracy}: The proportion of cases where the original variable name appears within the top-k results (typically top-10).

\item \textbf{Rank Distribution}: The frequency at which the correct name appears at each specific rank position from 1st through 10th.

\item \textbf{Mean Reciprocal Rank (MRR)}: Measures the average reciprocal rank position of the first correct prediction. Formally, given a set of queries $Q$, the MRR is calculated as:

$$
\text{MRR} = \frac{1}{|Q|} \sum_{i=1}^{|Q|} \frac{1}{\text{rank}_i}
$$

where $\text{rank}_i$ is the rank position of the first correct prediction for the $i$-th query. MRR emphasizes the importance of placing the first correct result as high as possible and is particularly suitable when only one relevant result is sufficient per query \cite{Mitr18a}.

\item \textbf{Normalized Discounted Cumulative Gain (NDCG)}: Measures the usefulness of predictions based on their positions and graded relevance. The DCG is calculated as:

$$
\text{DCG}_k = \sum_{i=1}^{k} \frac{2^{\text{rel}_i} - 1}{\log_2(i + 1)}
$$

where $\text{rel}_i$ represents the graded relevance of the prediction at position $i$. NDCG normalizes this score by dividing by the ideal DCG (IDCG), yielding:

$$
\text{NDCG}_k = \frac{\text{DCG}_k}{\text{IDCG}_k}
$$

thus accounting logarithmically for position discounting \cite{Mitr18a}.

\item \textbf{Timing and Memory Metrics}: Includes total and average completion times and memory usage, evaluated per prefix length.

\end{enumerate}

The benchmark can be run in various configurations, from baseline alphabetical sorters to advanced heuristic-guided engines. It can also be configured with different heuristic templates, offering flexibility to evaluate a wide range of completion strategies. Our benchmark is designed not only to test whether a completion system finds the correct result but also how efficiently it does so, reflecting real-world usage where users expect high precision with minimal typing. By measuring performance across multiple prefix lengths and analyzing rank-based metrics, we can assess the trade-offs and effectiveness of various completion algorithms. We chose MRR as our primary evaluation metric because it effectively captures the quality of ranked suggestions in code completion tasks. MRR reflects how early the correct suggestion appears in the list, aligning closely with real user experience. This makes it particularly suitable for comparing ranking-based approaches, such as the STAN-based reranking models we evaluate, and is consistent with evaluation protocols in prior work on neural code completion \cite{Svya21a}.
	
\subsection{Package Selection}
Based on the statistical analysis of Pharo code by Zaitsev \etal \cite{Zait20a} we selected packages that ensure a broad domain diversity, including web development, visualization, software analysis, and user-interface frameworks, while also focusing on projects that demonstrated active maintenance, extensive test suites, and representative structural attributes, such as method length distributions and language feature usage patterns. 

We selected Pharo packages reflecting significant diversity in domain, size, and development activity. The chosen packages span essential areas such as visualization (Roassal), software analysis (Moose), web application development (Seaside), user interface framework (Spec2), and version control systems (Iceberg). These packages were carefully selected based on their substantial adoption, active maintenance, extensive test coverage, and their use of key language features including polymorphism, reflective capabilities, and Pharo-specific syntax. This selection strategy ensures our benchmarks effectively capture common programming practices, typical complexity, and diverse usage patterns prevalent within the Pharo community. A comprehensive description of the selected packages is provided in Appendix~\ref{app:projects}.

\begin{table}[ht]
\centering
\begin{tabular}{llllllll}
\hline
Framework/Bib & \# Packages & \# Classes & \# Defined Classes & \# Methods & $\rho_{\text{int}}$ & $R_{\text{int}}$ & $R_{\text{ext}}$ \\
\hline
IceBerg & 11 & 610 & 525 & 4882 & 0.35 & 229 & 292 \\
\hline
Moose & 72 & 910 & 751 & 7581 & 0.21 & 221 & 1035 \\
\hline
Roassal & 52 & 794 & 629 & 7230 & 0.14 & 158 & 1311 \\
\hline
Seaside & 44 & 884 & 758 & 6605 & 0.21 & 305 & 552 \\
\hline
Spec & 33 & 1115 & 939 & 8383 & 0.21 & 278 & 593 \\
\hline
\end{tabular}
\caption{Overview of Selected Pharo Frameworks for Benchmark}
\label{bench}
\end{table}

Table \ref{bench} provides a comprehensive overview of the selected Pharo frameworks evaluated in our benchmarking experiments. It includes the total number of packages, classes, defined classes, and methods analyzed in each framework. Additionally, the table presents key metrics such as the ratio of internal references ($\rho_{\text{int}}$), number of internal references ($R_{\text{int}}$), and the number of external references ($R_{\text{ext}}$).  A higher $\rho_{\text{int}}$ indicates stronger intra-package cohesion, reflecting more frequent references to internal entities rather than external ones. For example, Iceberg shows a relatively high $\rho_{\text{int}}$ (0.35), suggesting significant internal cohesion, whereas Roassal exhibits the lowest $\rho_{\text{int}}$ (0.14), implying a greater reliance on external references. These metrics provide context for understanding how package structure and usage patterns influence the effectiveness of our package-aware heuristic.

\section{Results and Discussion}\label{discussion}

\subsection {Overall Results}

Our evaluation demonstrates the advantages and limitations of introducing a package-aware heuristic into Pharo's completion engine. Overall, we observed (Table \ref{bench2}) a significant improvement in MRR, indicating that developers receive more contextually relevant completion suggestions when package structure is leveraged. The average MRR improvement across all evaluated frameworks was notable, especially pronounced in frameworks like Spec (7.59\%) and Iceberg (6.09\%), which exhibit well-defined package structures with strong intra-package cohesion. However, the results also reveal nuanced behavior depending on the type of package and the framework's architectural characteristics. For instance, the Moose framework showed modest overall gains (1.05\%), which were more substantial in non-test packages (1.19\%) but negligible or negative in test packages (0.31\%). 

\begin{table}[ht]
\scriptsize
\centering
\begin{tabular}{lllllllllll}
\hline
Framework/Bib & Package Type & $\langle$ Metric $\rangle$ & MRR & 2 & 3 & 4 & 5 & 6 & 7 & 8 \\
\hline
\multirow{3}{*}{Iceberg}
  & \multirow{3}{*}{Overall} & Without & 31.27 & 5.36 & 16.91 & 27.82 & 37.45 & 39.91 & 45.00 & 49.91 \\
  &                           & With    & 37.36 & 12.64 & 23.82 & 37.18 & 44.00 & 43.45 & 49.73 & 54.64 \\
  &                           & $\Delta$&  6.09 &  7.27 &  6.91 &  9.36 &  6.55 &  3.55 &  4.73 &  4.73 \\ \cline{2-11}
\hline
\multirow{9}{*}{Moose}
  & \multirow{3}{*}{Overall} & Without & 40.53 & 17.32 & 19.98 & 31.92 & 37.95 & 44.31 & 49.92 & 58.95 \\
  &                           & With    & 41.58 & 23.19 & 23.47 & 31.44 & 37.15 & 44.37 & 49.32 & 58.44 \\
  &                           & $\Delta$&  1.05 &  5.86 &  3.49 & -0.47 & -0.80 &  0.07 & -0.59 & -0.51 \\ \cline{2-11}
  & \multirow{3}{*}{Test}  & Without & 22.09 & 3.79 & 10.29 & 17.93 & 21.57 & 26.43 & 32.36 & 47.71 \\
 & 				& With & 22.39 & 6.79 & 11.57 & 17.21 & 20.29 & 25.64 & 32.00 & 48.64 \\
  &                           & $\Delta$  & 0.31  & 3.00  & 1.29  & -0.71  & -1.29  & -0.79  & -0.36  & 0.93  \\\cline{2-11}
  & \multirow{3}{*}{Non-test}    & Without & 33.40 & 7.03 & 19.48 & 31.38 & 37.79 & 43.90 & 48.10 & 56.86 \\
 & 			     & With & 34.59 & 12.97 & 21.52 & 32.10 & 38.21 & 44.52 & 47.28 & 55.59 \\
  &                            & $\Delta$  & 1.19  & 5.93  & 2.03  & 0.72  & 0.41  & 0.62  & -0.83  & -1.28  \\
\hline
\multirow{9}{*}{Roassal}
  & \multirow{3}{*}{Overall} & Without & 34.74 &  3.16 & 15.40 & 29.94 & 43.68 & 51.04 & 56.86 & 65.10 \\
  &                           & With    & 35.64 &  6.00 & 18.54 & 31.48 & 45.34 & 50.98 & 54.02 & 63.10 \\
  &                           & $\Delta$&  0.90 &  2.84 &  3.14 &  1.54 &  1.66 & -0.06 & -2.84 & -2.00 \\ \cline{2-11}
  & \multirow{3}{*}{Test}   & Without & 32.38 & 1.00 & 9.69 & 26.31 & 42.46 & 51.00 & 56.92 & 62.77 \\
 & 				& With & 29.77 & 1.31 & 9.46 & 24.62 & 41.54 & 47.08 & 49.85 & 57.00 \\
  &                            & $\Delta$  & -2.62  & 0.31  & -0.23  & -1.69  & -0.92  & -3.92  & -7.08  & -5.77  \\ \cline{2-11}
  & \multirow{3}{*}{Non-test}    & Without & 35.57 & 3.92 & 17.41 & 31.22 & 44.11 & 51.05 & 56.84 & 65.92 \\
& 				 & With & 37.71 & 7.65 & 21.73 & 33.89 & 46.68 & 52.35 & 55.49 & 65.24 \\
  &                            & $\Delta$  & 2.14  & 3.73  & 4.32  & 2.68  & 2.57  & 1.30  & -1.35  & -0.68  \\
\hline
\multirow{9}{*}{Seaside}
  & \multirow{3}{*}{Overall} & Without & 37.72 &  6.28 & 16.75 & 32.38 & 43.53 & 51.28 & 62.25 & 66.94 \\
  &                           & With    & 42.44 &  9.41 & 24.38 & 37.66 & 49.16 & 55.03 & 65.59 & 69.94 \\
  &                           & $\Delta$&  4.72 &  3.13 &  7.62 &  5.28 &  5.63 &  3.75 &  3.34 &  3.00 \\ \cline{2-11}
  & \multirow{3}{*}{Test}    & Without & 34.55 & 7.27 & 17.27 & 31.45 & 41.18 & 47.09 & 56.82 & 58.09 \\
 & 				& With & 38.09 & 9.55 & 20.64 & 35.91 & 45.91 & 49.91 & 58.73 & 60.45 \\
  &                            & $\Delta$  & 3.55 & 2.27 & 3.36 & 4.45 & 4.73 & 2.82 & 1.91 & 2.36 \\ \cline{2-11}
  & \multirow{3}{*}{Non-test}  & Without & 39.38 & 5.76 & 16.48 & 32.86 & 44.76 & 53.48 & 65.10 & 71.57 \\
 & 				& With & 44.71 & 9.33 & 26.33 & 38.57 & 50.86 & 57.71 & 69.19 & 74.90 \\
  &                            & $\Delta$  & 5.33 & 3.57 & 9.86 & 5.71 & 6.10 & 4.24 & 4.10 & 3.33 \\
\hline
\multirow{9}{*}{Spec}
  & \multirow{3}{*}{Overall} & Without & 29.43 &  3.87 & 12.48 & 23.16 & 34.06 & 39.03 & 48.35 & 57.10 \\
  &                           & With    & 37.02 &  7.74 & 21.58 & 33.68 & 41.87 & 47.87 & 56.06 & 61.71 \\
  &                           & $\Delta$&  7.59 &  3.87 &  9.10 & 10.52 &  7.81 &  8.84 &  7.71 &  4.61 \\ \cline{2-11}
  & \multirow{3}{*}{Test}    & Without & 32.00 & 2.55 & 14.36 & 22.36 & 38.09 & 43.36 & 54.82 & 65.36 \\
 & 				& With & 34.64 & 4.55 & 19.36 & 28.73 & 40.82 & 46.18 & 56.36 & 61.91 \\
  &                            & $\Delta$  & 2.64  & 2.00  & 5.00  & 6.36  & 2.73  & 2.82  & 1.55  & -3.45  \\ \cline{2-11}
  & \multirow{3}{*}{Non-test}  & Without & 28.02 & 4.60 & 11.45 & 23.60 & 31.85 & 36.65 & 44.80 & 52.55 \\
 & 				&With & 38.32 & 9.50 & 22.80 & 36.40 & 42.45 & 48.80 & 55.90 & 61.60 \\
  &                            & $\Delta$  & 10.31  & 4.90  & 11.35  & 12.80  & 10.60  & 12.15  & 11.10  & 9.05  \\
\hline
\end{tabular}
\caption{MRR Performance Comparison between Default and Package-Aware across}
\label{bench2}
\end{table}

This highlights a key insight completion accuracy gains are context-sensitive, often dependent on package dependencies and naming conventions. Test packages, for example, frequently access numerous external global variables and classes, reducing the effectiveness of a strictly package-local prioritization heuristic. The Roassal framework exhibited similar mixed outcomes. Non-test packages show improvements (2.14\%), whereas test packages experienced performance degradation (-2.62\%). This negative outcome indicates that the current heuristic, which infers package relationships from naming conventions alone, may inadvertently prioritize less relevant local completions in testing scenarios, where global or cross-package dependencies are prevalent. 

Interestingly, Seaside and Spec demonstrated consistent improvements across both test and non-test packages, suggesting that in frameworks with strong internal modularization and delineated package structures, the heuristic significantly enhances completion relevance across diverse coding contexts. Another critical observation is that the heuristic's effectiveness diminishes as prefix lengths increase, with the most significant improvements consistently appearing at shorter prefix lengths (2-4 characters). This outcome aligns with real-world coding behavior, where developers rely heavily on early suggestions to reduce typing effort. However, the diminishing returns at longer prefixes suggest a reduced practical advantage once developers have provided extensive typing context. Our analysis also uncovered edge cases where similarly named global variables (\eg \ct{IceSBBrowserAbstractMethodCommand} vs. \ct{IceSBBrowseFullMethodCommand}) resulted in ambiguous completions. Addressing these cases may require additional heuristics or statistical models trained on historical usage data, highlighting an avenue for future work.

\subsection{Challenges}
In the appendix \ref{appendix}, detailed benchmarks for each evaluated framework are provided. It is important to note that, when computing averages and deltas presented in Table \ref{bench2}, zero values were omitted. A zero value typically represents cases where a package contains only one class with methods that are either trivial, purely for testing, or abstract (marked by \ct{self subclassResponsibility}) with concrete implementations residing outside the package. Additionally, some packages might only contain extensions without complete implementations.

The occurrence of unchanged or marginal delta values (\eg values close to 1) in MRR is often due to methods or variables sharing lengthy prefixes, such as \ct{IceSBBrowserAbstractMethodCommand} and \ct{IceSBBrowseFullMethodCommand}. Given that our current benchmarking evaluates prefixes limited to 8 characters, differences beyond this length remain undetected. This represents a significant limitation of our evaluation method.

To overcome this limitation, future research should explore more sophisticated completion strategies. For instance, completing only the common prefix and then progressively refining the completion. Consider an example from the Moose framework: typing the letter 'M' could immediately propose prefixes like 'Moose...', 'MooseMSE...', and 'MooseMSEImporter...'. Pressing 'Tab' after selecting a prefix would insert it directly without adding extra spaces, enabling seamless continued typing. Further keystrokes could then complete subsequent portions of the name (\eg, typing 'I' to complete 'Importer'). Such an approach could significantly reduce typing effort, allowing a long name like \ct{MooseMSEImporterTestEntity} to be entered with just a few keystrokes ('M' - 'Tab' completes 'MooseMSE', then 'I' - 'Tab' completes 'ImporterTestEntity'). Integrating this prefix-driven completion strategy into future benchmarks will provide deeper insights into code completion effectiveness, especially in projects characterized by extensive naming conventions.

\section{Limitations} \label{limitations}
A key limitation of our evaluation is its dependence on static references extracted from existing code. We simulate completion sites by truncating identifier names to 2–8 characters and then measuring whether our approach ranks the correct name near the top of the suggestion list. Although this technique is standard in code-completion research, it does not fully capture how developers behave in live sessions. Results could also differ when analyzing older repositories or external libraries that diverge from Pharo’s usual package conventions.

Additionally, an important assumption underlying our heuristics is that global variables or classes are more likely to be referenced within the same package where they are defined, rather than in other packages within the same project. This assumption justifies our strategy of prioritizing classes first from the same package, followed by "friend" packages, and finally external packages. However, this assumption has not been empirically validated, and there are multiple scenarios where it may not hold true. For instance, test packages rarely reference classes from the same package but frequently reference classes from other packages within the same project. Similarly, classes defined in core packages or packages implementing design patterns such as visitors, exceptions, or commands often exhibit fewer references within their own packages but are extensively referenced by others.

To address this potential threat to validity, future work should investigate the actual degree of self-referencing within packages. If self-referencing is found to be low, prioritizing classes from the same package could negatively impact completion accuracy, as indicated by our results in Table \ref{bench2} for test packages. Thus, it might be beneficial to prioritize classes from the same project rather than strictly from the same package, potentially improving completion suggestions, particularly in cases such as test packages.

Finally, very long prefixes, such as \ct{IceSBBrowserAbstractMethodCommand} and \ct{IceSBBrowseFullMethodCommand} remain challenging because our current methodology considers only short prefixes. Future work should extend the analysis to longer prefixes and a broader range of code bases, thereby providing a more complete picture of completion behavior and ultimately yielding more accurate, context-aware suggestions for developers.

\section{Related Works} \label{relatedwork}
Code completion has evolved from simple syntactic suggestions to sophisticated, context-sensitive systems powered by statistical, neural, and usability-aware methods. Early work such as Robbes \etal \cite{Robb08a} introduced context-sensitive filtering based on recent usage and program history. Shortly afterward, Bruch \etal \cite{Bruc09a} and Hou \etal \cite{Hou10a} emphasized syntactic similarity and ranking heuristics, relying on type hierarchies, usage popularity, and manual grouping. These heuristic and rule-based approaches established foundational techniques for example-based ranking and structural filtering still used in IDE plugins today.

\paragraph{Statistical Approaches}  Hindle \etal \cite{Hind12a} demonstrated that software code exhibits significant regularities that can be effectively captured using statistical language models, specifically through n-gram modeling. Statistical methods soon emerged, bringing greater accuracy through learning from large corpora. Nguyen \etal \cite{Nguy13a} introduced SLAMC, combining n-gram models with semantic roles and topic modeling. Raychev \etal contributed SLANG \cite{Rayc14a}, a statistical language model synthesizing code completions. Proksch \etal \cite{Prok15a} introduced Bayesian models offering compact and accurate recommendations via probabilistic reasoning over API usage patterns. Nguyen \etal \cite{Nguy16a} further advanced statistical methods with APIREC, learning fine-grained API usage patterns. Raychev \etal subsequently presented DEEP3 \cite{Raych16a}, employing decision-tree-based generative models.

\paragraph{Learning-Based Approaches} Deep neural architectures were increasingly adopted for code completion. Bielik \etal \cite{Biel16a} proposed PHOG, a grammar-aware generative model using probabilistic higher-order rules. Jin \etal \cite{Jin18a} highlighted usability concerns by addressing the hidden costs of extensive suggestion lists. Hellendoorn \etal \cite{Hell19a} provided a significant empirical analysis revealing the limitations of models in practical intra-project completions. Svyatkovskiy \etal \cite{Svya21a} introduced a modular neural framework for code completion, leveraging static analysis and granular token encodings to design a memory-efficient reranking model with high predictive performance. Karampatsis \etal \cite{Kara20a} introduced open-vocabulary neural models with Byte-Pair Encoding (BPE), effectively managing out-of-vocabulary issues. Matani \etal \cite{Mata21a} proposed an efficient segment-tree-based solution for prefix-based completion without the need for statistical training. Popov \etal \cite{Popo21a} delivered a practical, time-efficient GPT-2 variant for R code completion, achieving high accuracy within strict latency constraints. Li \etal \cite{Li21b} proposed benefit-cost-aware metrics to filter and reorder completions, significantly reducing irrelevant suggestions. 

\paragraph{Recent contributions} Recent contributions emphasize real-time usability, contextual awareness, and cross-language generalization. Bibaev \etal \cite{Biba22a} advanced towards practical systems by training rankers using real IDE usage logs, and personalizing suggestions to reduce developer keystroke effort. Takerngsaksiri \etal \cite{Take24a} introduced PyCoder, a syntax-aware transformer-based model predicting token types without explicit AST parsing. Wang \etal \cite{Wang25b} proposed TIGER, a generate-then-rank approach using lightweight transformers for Python type inference, demonstrating state-of-the-art performance. Modern code completion systems now integrate structural analysis, statistical learning, and deep neural modeling, consistently targeting developer productivity through usability-driven metrics. Despite these advancements, significant challenges remain, especially in handling difficult intra-project completions and minimizing cognitive load. Future developments must balance sophistication, runtime efficiency, and developer experience, increasingly turning to hybrid models and log-informed personalization.

\section{Conclusion and future works} \label{conclusion}

\compl employs multiple semantic heuristics to produce context-aware completions in a live programming environment. However, it originally treated the system as a flat global space, limiting the precision of suggestions for large projects. In this paper, we proposed and evaluated a package-awareness completion heuristic to mitigate this issue. Our approach prioritizes local package entities, followed by similarly prefixed or related packages, and finally falls back to the global namespace. The results show that package-awareness completions indeed improve the ranking of relevant suggestions in some cases, especially when a package’s references remain largely local. Conversely, packages with major cross-package dependencies, particularly testing packages, can perform worse with naive prefix-based ordering, highlighting the need to explicitly consider package dependencies and usage patterns.

We plan to integrate more advanced dependency-aware scoping, so that \compl understands not just one’s immediate package but also any dependencies or related packages. We will also explore hybrid approaches that combine lightweight statistical frequency analysis with structural heuristics to handle complex referencing patterns. By evolving in this direction, we aim to make \compl an increasingly effective code-completion system for large, modular software projects.

\paragraph{Acknowledgments.}
We thank Inria and the LLM4Code défi for the funding of the first author. 
\bibliographystyle{alpha}
\bibliography{rmod,others}

\newcommand{\etalchar}[1]{$^{#1}$}
\begin{thebibliography}{DHDJML24}

\bibitem[AEH{\etalchar{+}}20]{Anqu20a}
Nicolas Anquetil, Anne Etien, Mahugnon~Honor\'e Houekpetodji, Beno{\^\i}t
  Verhaeghe, St{\'e}phane Ducasse, Clotilde Toullec, Fatija Djareddir,
  J\'er\^ome Sudich, and Mustapha Derras.
\newblock Modular {Moose}: A new generation of software reengineering platform.
\newblock In {\em International Conference on Software and Systems Reuse
  (ICSR'20)}, number 12541 in LNCS, pages 119--134, December 2020.

\bibitem[BDN{\etalchar{+}}09]{Blac09a}
Andrew~P. Black, St{\'e}phane Ducasse, Oscar Nierstrasz, Damien Pollet, Damien
  Cassou, and Marcus Denker.
\newblock {\em Pharo by Example}.
\newblock Square Bracket Associates, Kehrsatz, Switzerland, 2009.

\bibitem[BDR08]{Berg08a}
Alexandre Bergel, St{\'e}phane Ducasse, and Lukas Renggli.
\newblock Seaside -- advanced composition and control flow for dynamic web
  applications.
\newblock {\em ERCIM News}, 72, January 2008.

\bibitem[Ber16]{Berg16c}
Alexandre Bergel.
\newblock {\em Agile Visualization}.
\newblock LULU Press, 2016.

\bibitem[BKL{\etalchar{+}}22]{Biba22a}
Vitaliy Bibaev, Alexey Kalina, Vadim Lomshakov, Yaroslav Golubev, Alexander
  Bezzubov, Nikita Povarov, and Timofey Bryksin.
\newblock All you need is logs: Improving code completion by learning from
  anonymous {IDE} usage logs, 2022.

\bibitem[BMM09]{Bruc09a}
Marcel Bruch, Martin Monperrus, and Mira Mezini.
\newblock Learning from examples to improve code completion systems.
\newblock In {\em Proceedings of the 7th joint meeting of the European software
  engineering conference and the ACM SIGSOFT symposium on the foundations of
  software engineering}, pages 213--222, 2009.

\bibitem[BRV16]{Biel16a}
Pavol Bielik, Veselin Raychev, and Martin Vechev.
\newblock {PHOG}: Probabilistic model for code.
\newblock In {\em International Conference on Machine Learning}, 2016.

\bibitem[CNG15]{Chis15b}
Andrei Chi\c{s}, Oscar Nierstrasz, and Tudor G\^{\i}rba.
\newblock Towards moldable development tools.
\newblock In {\em Workshop on Evaluation and Usability of Programming Languages
  and Tools}, 2015.

\bibitem[DHDJML24]{DeHon24a}
Koen De~Hondt, S{\'e}phane Ducasse, Sebastian Jordan~Montano, and Esteban
  Lorenzano.
\newblock {\em Application Building with Spec 2.0}.
\newblock Book on Demand -- Keepers of the lighthouse, 2024.

\bibitem[DLR07]{Duca07a}
St{\'e}phane Ducasse, Adrian Lienhard, and Lukas Renggli.
\newblock Seaside: A flexible environment for building dynamic web
  applications.
\newblock {\em IEEE Software}, 24(5):56--63, 2007.

\bibitem[DRS{\etalchar{+}}10]{Duca10a}
St{\'e}phane Ducasse, Lukas Renggli, C.~David Shaffer, Rick Zaccone, and
  Michael Davies.
\newblock {\em Dynamic Web Development with Seaside}.
\newblock Square Bracket Associates, 2010.

\bibitem[GHJV95]{Gamm95a}
Erich Gamma, Richard Helm, Ralph Johnson, and John Vlissides.
\newblock {\em Design Patterns: Elements of Reusable Object-Oriented Software}.
\newblock Addison-Wesley, 1995.

\bibitem[HBS{\etalchar{+}}12]{Hind12a}
Abram Hindle, Earl~T Barr, Zhendong Su, Mark Gabel, and Premkumar Devanbu.
\newblock On the naturalness of software.
\newblock In {\em Software Engineering ({ICSE}), 2012 34th International
  Conference on}, pages 837--847. IEEE, 2012.

\bibitem[HP10]{Hou10a}
Daqing Hou and David~M. Pletcher.
\newblock Towards a better code completion system by api grouping, filtering,
  and popularity-based ranking.
\newblock In {\em International Workshop on Recommendation Systems for Software
  Engineering (RSSE)}, 2010.

\bibitem[HPGB19]{Hell19a}
Vincent~J Hellendoorn, Sebastian Proksch, Harald~C Gall, and Alberto Bacchelli.
\newblock When code completion fails: A case study on real-world completions.
\newblock In {\em 2019 IEEE/ACM 41st International Conference on Software
  Engineering ({ICSE})}, pages 960--970. IEEE, 2019.

\bibitem[JH25]{Jin25a}
Hangzhan Jin and Mohammad Hamdaqa.
\newblock Ccci: Code completion with contextual information for complex data
  transfer tasks using large language models.
\newblock {\em arXiv}, 2025.

\bibitem[JS18]{Jin18a}
Xianhao Jin and Francisco Servant.
\newblock The hidden cost of code completion: understanding the impact of the
  recommendation-list length on its efficiency.
\newblock In {\em International Conference on Mining Software Repositories},
  2018.

\bibitem[KBR{\etalchar{+}}20]{Kara20a}
Rafael-Michael Karampatsis, Hlib Babii, Romain Robbes, Charles Sutton, and
  Andrea Janes.
\newblock Big code != big vocabulary: open-vocabulary models for source code.
\newblock In {\em International Conference on Software Engineering (ICSE)},
  2020.

\bibitem[LHL{\etalchar{+}}21]{Li21b}
Jingxuan Li, Rui Huang, Wei Li, Kai Yao, and Weiguo Tan.
\newblock {Toward Less Hidden Cost of Code Completion with Acceptance and
  Ranking Models}.
\newblock In {\em International Conference on Software Maintenance and
  Evolution (ICSME)}, 2021.

\bibitem[Mat21]{Mata21a}
Dhruv Matani.
\newblock An \$o(k {\textbackslash}log\{n\})\$ algorithm for prefix based
  ranked autocomplete.
\newblock {\em {arXiv}}, 2021.

\bibitem[MC18]{Mitr18a}
Bhaskar Miutra and Nick Craswell.
\newblock An introduction to neural information retrieval.
\newblock {\em Foundations and Trends in Information Retrieval}, 2018.

\bibitem[NDG05]{Nier05c}
Oscar Nierstrasz, St{\'e}phane Ducasse, and Tudor G\^irba.
\newblock The story of {Moose}: an agile reengineering environment.
\newblock In Michel Wermelinger and Harald Gall, editors, {\em Proceedings of
  the European Software Engineering Conference}, ESEC/FSE'05, pages 1--10, New
  York NY, 2005. ACM Press.
\newblock Invited paper.

\bibitem[NHC{\etalchar{+}}16]{Nguy16a}
Anh~Tuan Nguyen, Michael Hilton, Mihai Codoban, Hoan~Anh Nguyen, Lily Mast, Eli
  Rademacher, Tien~N. Nguyen, and Danny Dig.
\newblock Api code recommendation using statistical learning from fine-grained
  changes.
\newblock In {\em International Symposium on Foundations of Software
  Engineering}, 2016.

\bibitem[NNNN13]{Nguy13a}
Tung~Thanh Nguyen, Anh~Tuan Nguyen, Hoan~Anh Nguyen, and Tien~N Nguyen.
\newblock A statistical semantic language model for source code.
\newblock In {\em Proceedings of the 2013 9th Joint Meeting on Foundations of
  Software Engineering}, pages 532--542, 2013.

\bibitem[PDO20]{Poli20y}
Guillermo Polito, St{\'e}phane Ducasse, and Allex Oliveira.
\newblock Manage your code with git and iceberg, 2020.

\bibitem[PLM15]{Prok15a}
Sebastian Proksch, Johannes Lerch, and Mira Mezini.
\newblock Intelligent code completion with bayesian networks.
\newblock {\em Transactions on Software Engineering and Methodology (TOSEM)},
  1(25), 2015.

\bibitem[POL{\etalchar{+}}21]{Popo21a}
Artem Popov, Dmitrii Orekhov, Denis Litvinov, Nikolay Korolev, and Gleb
  Morgachev.
\newblock Time-efficient code completion model for the r programming language.
\newblock In {\em Workshop on Natural Language Processing for Programming},
  2021.

\bibitem[RBV16]{Raych16a}
Veselin Raychev, Pavol Bielik, and Martin Vechev.
\newblock Probabilistic model for code with decision trees.
\newblock In {\em International Conference on Object-Oriented Programming,
  Systems, Languages, and Applications}, 2016.

\bibitem[RL08a]{Robb08a}
Romain Robbes and Michele Lanza.
\newblock How program history can improve code completion.
\newblock In {\em Proceedings of ASE 2008 (23rd International Conference on
  Automated Software Engineering)}, pages 317--326, 2008.

\bibitem[RL08b]{Robb08b}
Romain Robbes and Michele Lanza.
\newblock Spy{W}are: a change-aware development toolset.
\newblock In {\em Proceedings of the 30th International Conference on Software
  Engineering}, ICSE'08, pages 847--850, New York, NY, USA, 2008. ACM.

\bibitem[RL10]{Robb10c}
Romain Robbes and Michele Lanza.
\newblock Improving code completion with program history.
\newblock {\em Journal of Automated Software Engineering}, 17(2):181--212,
  2010.

\bibitem[RVY14]{Rayc14a}
Veselin Raychev, Martin Vechev, and Eran Yahav.
\newblock Code completion with statistical language models.
\newblock In {\em Acm Sigplan Notices}, volume~49, pages 419--428. ACM, 2014.

\bibitem[SLH{\etalchar{+}}21]{Svya21a}
Alexey Svyatkovskiy, Sebastian Lee, Anna Hadjitofi, Maik Riechert, Juliana
  Franco, and Miltiadis Allamanis.
\newblock Fast and memory-efficient neural code completion.
\newblock In {\em International Conference on Mining Software Repositories
  (MSR)}, 2021.

\bibitem[TTL24]{Take24a}
Wannita Takerngsaksiri, Chakkrit Tantithamthavorn, and Yuan-Fang Li.
\newblock Syntax-aware on-the-fly code completion.
\newblock {\em Information and Software Technology}, 2024.

\bibitem[WZL{\etalchar{+}}25]{Wang25b}
Chong Wang, Jian Zhang, Yiling Lou, Mingwei Liu, Weisong Sun, Yang Liu, and Xin
  Peng.
\newblock {TIGER: A Generating-Then-Ranking Framework for Practical Python Type
  Inference }.
\newblock In {\em International Conference on Software Engineering (ICSE)},
  2025.

\bibitem[ZDA20]{Zait20a}
Oleksandr Zaitsev, St{\'e}phane Ducasse, and Nicolas Anquetil.
\newblock Characterizing pharo code: A technical report.
\newblock Technical report, Inria Lille Nord Europe - Laboratoire CRIStAL -
  Universit\'e de Lille ; Arolla, January 2020.

\end{thebibliography}

\newpage
\appendix \label{appendix}

\newpage
\section{Project Descriptions}
\label{app:projects}

\paragraph{Spec.}
Spec is a modern user interface framework integrated into Pharo. It adopts a modular architecture centered around "presenters," which allows developers to efficiently compose, nest, and manage interactive UI elements \cite{DeHon24a}. Spec2 facilitates dynamic interface layouts, enabling real-time modifications without the need for extensive interface rebuilds, thus enhancing adaptability and responsiveness. Additionally, it supports multiple rendering backends, including Morphic and GTK+3, providing flexibility for cross-platform application development. Spec2's design promotes streamlined communication between components, significantly simplifying interaction handling and improving maintainability. 

\paragraph{Roassal.} Roassal is a lightweight and extensible visualization engine developed in Pharo, designed to facilitate agile and interactive data visualization \cite{Berg16c}. It provides a rich set of graphical primitives and layout algorithms, enabling developers to craft domain-specific visualizations with minimal effort. Roassal supports interactive features such as zooming, dragging, and tooltips, allowing users to explore complex data structures dynamically. Its integration with the Pharo environment allows for seamless development and immediate feedback, making it an effective tool for both exploratory data analysis and the development of custom visualization tools. 

\paragraph{Seaside.}
Seaside is a web application framework for Smalltalk, particularly well-integrated with Pharo, that enables the development of complex web applications through a component-based architecture \cite{Duca07a, Duca10a}. Unlike traditional web frameworks that rely on templates, Seaside allows developers to build web pages by composing stateful components, each encapsulating its rendering and behavior. It leverages continuations to manage control flow, facilitating the creation of sophisticated user interactions and workflows. Seaside's approach promotes code reuse and modularity, and its tight integration with the Pharo development environment allows for live debugging and real-time updates, enhancing developer productivity \cite{Berg08a}.  

\paragraph{Iceberg.}
Iceberg is the primary version control system (VCS) integration tool within the Pharo environment, providing a seamless interface to Git repositories. It enables developers to perform standard Git operations such as cloning, committing, branching, merging, and pushing directly from the Pharo image, eliminating the need for external command-line tools. Designed to bridge the gap between Pharo's live object model and Git's file-based architecture, Iceberg ensures that changes made within the Pharo environment are accurately reflected in the Git repository. This integration facilitates efficient management of code versions, supports collaborative development workflows, and simplifies the process of contributing to and maintaining Pharo-based projects \cite{Poli20y}. 

\paragraph{Moose.}
Moose is an open-source platform for software and data analysis, developed in Pharo, that enables analysts to construct custom analysis tools and workflows \cite{Anqu20a}. It offers services such as data importation, modeling, measurement, querying, mining, and the development of interactive visual analysis tools.  Moose supports the creation of dedicated analysis tools and the customization of analysis processes. It provides mechanisms for importing and meta-modeling through a generic meta-described engine, parsing using various technologies, and visualization via graph and chart engines.  The platform is primarily used for software analysis but is designed to handle various types of data. Moose is based on Pharo and is open-source under BSD/MIT licenses \cite{Nier05c}.

\newpage
\section{Seaside}
\scriptsize
\begin{longtable}{llllllllll}
  \caption{Performance of Seaside Framework }\\
\hline
Package & Metric & MRR & 2 & 3 & 4 & 5 & 6 & 7 & 8 \\
\hline
\endfirsthead

\multicolumn{10}{c}%
{{\bfseries \tablename\ \thetable{} -- continued from previous page}} \\
\hline
Package & Metric & MRR & 2 & 3 & 4 & 5 & 6 & 7 & 8 \\
\hline
\endhead

\hline \multicolumn{10}{r}{{Continued on next page}} \\
\endfoot

\hline
\endlastfoot
Seaside-Ajaxifier-Core & without & 0 & 0.0 & 0.0 & 0.0 & 0.0 & 0.0 & 0.0 & 0.0 \\
Seaside-Ajaxifier-Core & with & 0 & 0.0 & 0.0 & 0.0 & 0.0 & 0.0 & 0.0 & 0.0 \\
\hline
Seaside-Canvas & without & 0.48 & 0.04 & 0.17 & 0.44 & 0.6 & 0.67 & 0.72 & 0.8 \\
Seaside-Canvas & with & 0.55 & 0.05 & 0.31 & 0.57 & 0.68 & 0.72 & 0.77 & 0.85 \\
\hline
Seaside-Component & without & 0.40& 0.02 & 0.12 & 0.39 & 0.48 & 0.58 & 0.64 & 0.74 \\
Seaside-Component & with & 0.60 & 0.16 & 0.5 & 0.62 & 0.69 & 0.77 & 0.77 & 0.85 \\
\hline
Seaside-Continuation & without & 0 & 0.0 & 0.0 & 0.0 & 0.0 & 0.0 & 0.0 & 0.0 \\
Seaside-Continuation & with & 0 & 0.0 & 0.0 & 0.0 & 0.0 & 0.0 & 0.0 & 0.0 \\
\hline
Seaside-Core & without & 0.31& 0.03 & 0.12 & 0.29 & 0.4 & 0.43 & 0.49 & 0.53 \\
Seaside-Core & with & 0.42 & 0.04 & 0.24 & 0.43 & 0.53 & 0.55 & 0.63 & 0.67 \\
\hline
Seaside-Development & without & 0.36 & 0.02 & 0.09 & 0.27 & 0.41 & 0.45 & 0.67 & 0.75 \\
Seaside-Development & with & 0.45 & 0.08 & 0.28 & 0.41 & 0.51 & 0.53 & 0.72 & 0.79 \\
\hline
Seaside-Email & without & 0.29 & 0.05 & 0.15 & 0.26 & 0.29 & 0.38 & 0.46 & 0.52 \\
Seaside-Email & with & 0.37 & 0.08 & 0.3 & 0.41 & 0.42 & 0.42 & 0.5 & 0.54 \\
\hline
Seaside-Environment & without & 0.26& 0.0 & 0.0 & 0.06 & 0.1 & 0.1 & 0.79 & 0.79 \\
Seaside-Environment & with & 0.34& 0.0 & 0.07 & 0.22 & 0.22 & 0.22 & 0.85 & 0.85 \\
\hline
Seaside-Examples & without & 0.32 & 0.03 & 0.1 & 0.21 & 0.36 & 0.38 & 0.66 & 0.69 \\
Seaside-Examples & with & 0.38 & 0.07 & 0.29 & 0.33 & 0.36 & 0.37 & 0.67 & 0.71 \\
\hline
Seaside-Flow & without & 0 & 0.0 & 0.0 & 0.0 & 0.0 & 0.0 & 0.0 & 0.0 \\
Seaside-Flow & with & 0 & 0.0 & 0.0 & 0.0 & 0.0 & 0.0 & 0.0 & 0.0 \\
\hline
Seaside-JSON-Core & without & 0.35 & 0.09 & 0.22 & 0.33 & 0.35 & 0.36 & 0.61 & 0.61 \\
Seaside-JSON-Core & with & 0.34 & 0.1 & 0.2 & 0.31 & 0.33 & 0.34 & 0.61 & 0.61 \\
\hline
Seaside-Pharo-Canvas & without & 0 & 0.0 & 0.0 & 0.0 & 0.0 & 0.0 & 0.0 & 0.0 \\
Seaside-Pharo-Canvas & with & 0 & 0.0 & 0.0 & 0.0 & 0.0 & 0.0 & 0.0 & 0.0 \\
\hline
Seaside-Pharo-Continuation & without & 0.40 & 0.07 & 0.32 & 0.38 & 0.45 & 0.58 & 0.58 & 0.58 \\
Seaside-Pharo-Continuation & with & 0.44 & 0.22 & 0.38 & 0.4 & 0.45 & 0.58 & 0.58 & 0.58 \\
\hline
Seaside-Pharo-Core & without & 0.53 & 0.14 & 0.14 & 0.14 & 0.33 & 1.0 & 1.0 & 1.0 \\
Seaside-Pharo-Core & with & 0.53 & 0.14 & 0.14 & 0.14 & 0.33 & 1.0 & 1.0 & 1.0 \\
\hline
Seaside-Pharo-Development & without & 0.43 & 0.08 & 0.13 & 0.18 & 0.59 & 0.64 & 0.74 & 0.83 \\
Seaside-Pharo-Development & with & 0.42 & 0.09 & 0.16 & 0.19 & 0.58 & 0.62 & 0.71 & 0.79 \\
\hline
Seaside-Pharo-Email & without & 0.28 & 0.0 & 0.0 & 0.13 & 0.21 & 0.43 & 0.57 & 1.0 \\
Seaside-Pharo-Email & with & 0.33 & 0.0 & 0.0 & 0.15 & 0.45 & 0.45 & 0.6 & 1.0 \\
\hline
Seaside-Pharo-Environment & without & 0 & 0.0 & 0.0 & 0.0 & 0.0 & 0.0 & 0.0 & 0.0 \\
Seaside-Pharo-Environment & with & 0 & 0.0 & 0.0 & 0.0 & 0.0 & 0.0 & 0.0 & 0.0 \\
\hline
Seaside-Pharo-Flow & without & 0 & 0.0 & 0.0 & 0.0 & 0.0 & 0.0 & 0.0 & 0.0 \\
Seaside-Pharo-Flow & with & 0 & 0.0 & 0.0 & 0.0 & 0.0 & 0.0 & 0.0 & 0.0 \\
\hline
Seaside-Pharo-JSON-Core & without & 0 & 0.0 & 0.0 & 0.0 & 0.0 & 0.0 & 0.0 & 0.0 \\
Seaside-Pharo-JSON-Core & with & 0 & 0.0 & 0.0 & 0.0 & 0.0 & 0.0 & 0.0 & 0.0 \\
\hline
Seaside-Pharo-Tools-Web & without & 0.52 & 0.22 & 0.41 & 0.49 & 0.57 & 0.64 & 0.69 & 0.7 \\
Seaside-Pharo-Tools-Web & with & 0.53 & 0.22 & 0.44 & 0.51 & 0.59 & 0.65 & 0.71 & 0.72 \\
\hline
Seaside-Pharo-Welcome & without & 0 & 0.0 & 0.0 & 0.0 & 0.0 & 0.0 & 0.0 & 0.0 \\
Seaside-Pharo-Welcome & with & 0 & 0.0 & 0.0 & 0.0 & 0.0 & 0.0 & 0.0 & 0.0 \\
\hline
Seaside-Pharo100-Tools-Spec2 & without & 0.49 & 0.03 & 0.27 & 0.42 & 0.68 & 0.68 & 0.68 & 0.76 \\
Seaside-Pharo100-Tools-Spec2 & with & 0.50 & 0.03 & 0.27 & 0.34 & 0.72 & 0.72 & 0.72 & 0.81 \\
\hline
Seaside-Pharo90-REST-Core & without & 0 & 0.0 & 0.0 & 0.0 & 0.0 & 0.0 & 0.0 & 0.0 \\
Seaside-Pharo90-REST-Core & with & 0 & 0.0 & 0.0 & 0.0 & 0.0 & 0.0 & 0.0 & 0.0 \\
\hline
Seaside-REST-Core & without & 0.50 & 0.1 & 0.23 & 0.52 & 0.62 & 0.65 & 0.67 & 0.76 \\
Seaside-REST-Core & with & 0.60& 0.19 & 0.53 & 0.68 & 0.68 & 0.69 & 0.72 & 0.79 \\
\hline
Seaside-RenderLoop & without & 0.43 & 0.01 & 0.14 & 0.43 & 0.54 & 0.62 & 0.68 & 0.7 \\
Seaside-RenderLoop & with & 0.44 & 0.06 & 0.2 & 0.36 & 0.49 & 0.57 & 0.72 & 0.75 \\
\hline
Seaside-Session & without & 0.39 & 0.01 & 0.14 & 0.4 & 0.55 & 0.56 & 0.56 & 0.6 \\
Seaside-Session & with & 0.45 & 0.02 & 0.17 & 0.44 & 0.6 & 0.63 & 0.67 & 0.7 \\
\hline
Seaside-Tests-Canvas & without & 0.25 & 0.02 & 0.07 & 0.13 & 0.22 & 0.23 & 0.64 & 0.68 \\
Seaside-Tests-Canvas & with & 0.30 & 0.03 & 0.14 & 0.2 & 0.29 & 0.29 & 0.64 & 0.69 \\
\hline
Seaside-Tests-Component & without & 0.38 & 0.03 & 0.15 & 0.32 & 0.43 & 0.62 & 0.69 & 0.73 \\
Seaside-Tests-Component & with & 0.38 & 0.1 & 0.23 & 0.33 & 0.43 & 0.52 & 0.61 & 0.66 \\
\hline
Seaside-Tests-Core & without & 0.27 & 0.02 & 0.1 & 0.24 & 0.37 & 0.38 & 0.45 & 0.46 \\
Seaside-Tests-Core & with & 0.27 & 0.02 & 0.09 & 0.26 & 0.35 & 0.38 & 0.45 & 0.46 \\
\hline
Seaside-Tests-Environment & without & 0.32 & 0.01 & 0.09 & 0.31 & 0.42 & 0.44 & 0.56 & 0.5 \\
Seaside-Tests-Environment & with & 0.38 & 0.01 & 0.15 & 0.35 & 0.48 & 0.55 & 0.67 & 0.62 \\
\hline
Seaside-Tests-Flow & without & 0.40 & 0.02 & 0.13 & 0.46 & 0.53 & 0.57 & 0.61 & 0.6 \\
Seaside-Tests-Flow & with & 0.39 & 0.04 & 0.13 & 0.35 & 0.54 & 0.58 & 0.61 & 0.61 \\
\hline
Seaside-Tests-Functional & without & 0.37 & 0.1 & 0.2 & 0.32 & 0.44 & 0.48 & 0.58 & 0.7 \\
Seaside-Tests-Functional & with & 0.39 & 0.1 & 0.24 & 0.33 & 0.44 & 0.48 & 0.6 & 0.7 \\
\hline
Seaside-Tests-Pharo-Canvas & without & 0 & 0.0 & 0.0 & 0.0 & 0.0 & 0.0 & 0.0 & 0.0 \\
Seaside-Tests-Pharo-Canvas & with & 0 & 0.0 & 0.0 & 0.0 & 0.0 & 0.0 & 0.0 & 0.0 \\
\hline
Seaside-Tests-Pharo-Continuation & without & 0.46 & 0.14 & 0.42 & 0.44 & 0.45 & 0.71 & 0.75 & 0.75 \\
Seaside-Tests-Pharo-Continuation & with & 0.32 & 0.14 & 0.22 & 0.29 & 0.4 & 0.5 & 0.5 & 0.5 \\
\hline
Seaside-Tests-Pharo-Core & without & 0.32& 0.12 & 0.19 & 0.29 & 0.37 & 0.38 & 0.48 & 0.55 \\
Seaside-Tests-Pharo-Core & with & 0.32 & 0.12 & 0.17 & 0.3 & 0.37 & 0.38 & 0.47 & 0.55 \\
\hline
Seaside-Tests-Pharo-Functional & without & 0.46& 0.33 & 0.37 & 0.51 & 0.51 & 0.51 & 0.51 & 0.51 \\
Seaside-Tests-Pharo-Functional & with & 0.69 & 0.33 & 0.38 & 0.83 & 0.83 & 0.83 & 0.83 & 0.83 \\
\hline
Seaside-Tests-RenderLoop & without & 0.26 & 0.0 & 0.03 & 0.22 & 0.35 & 0.37 & 0.49 & 0.42 \\
Seaside-Tests-RenderLoop & with & 0.49 & 0.15 & 0.4 & 0.52 & 0.56 & 0.56 & 0.66 & 0.61 \\
\hline
Seaside-Tests-Session & without & 0.31 & 0.01 & 0.15 & 0.22 & 0.44 & 0.49 & 0.49 & 0.49 \\
Seaside-Tests-Session & with & 0.26 & 0.01 & 0.12 & 0.19 & 0.36 & 0.42 & 0.42 & 0.42 \\
\hline
Seaside-Tools-Core & without & 0 & 0.0 & 0.0 & 0.0 & 0.0 & 0.0 & 0.0 & 0.0 \\
Seaside-Tools-Core & with & 0 & 0.0 & 0.0 & 0.0 & 0.0 & 0.0 & 0.0 & 0.0 \\
\hline
Seaside-Tools-Web & without & 0.40 & 0.04 & 0.13 & 0.35 & 0.49 & 0.58 & 0.68 & 0.72 \\
Seaside-Tools-Web & with & 0.49 & 0.08 & 0.3 & 0.45 & 0.59 & 0.64 & 0.73 & 0.76 \\
\hline
Seaside-Welcome & without & 0.38 & 0.04 & 0.15 & 0.47 & 0.5 & 0.53 & 0.58 & 0.5 \\
Seaside-Welcome & with & 0.35 & 0.09 & 0.16 & 0.32 & 0.49 & 0.51 & 0.54 & 0.45 \\
\hline
Seaside-Widgets & without & 0.36 & 0.17 & 0.26 & 0.38 & 0.36 & 0.43 & 0.57 & 0.81 \\
Seaside-Widgets & with & 0.44 & 0.19 & 0.37 & 0.42 & 0.43 & 0.58 & 0.68 & 0.87 \\
\hline
Seaside-Zinc-Core & without & 0.39 & 0.02 & 0.17 & 0.36 & 0.52 & 0.54 & 0.63 & 0.64 \\
Seaside-Zinc-Core & with & 0.42 & 0.05 & 0.22 & 0.4 & 0.54 & 0.56 & 0.63 & 0.64 \\
\hline
Seaside-Zinc-Pharo & without & 0 & 0.0 & 0.0 & 0.0 & 0.0 & 0.0 & 0.0 & 0.0 \\
Seaside-Zinc-Pharo & with & 0 & 0.0 & 0.0 & 0.0 & 0.0 & 0.0 & 0.0 & 0.0 \\
\end{longtable}

\newpage

\section{Spec}
\scriptsize
\begin{longtable}{llllllllll}
  \caption{Performance of Spec Framework }\\
\hline
Package & Metric & MRR & 2 & 3 & 4 & 5 & 6 & 7 & 8 \\
\hline
\endfirsthead

\multicolumn{10}{c}%
{{\bfseries \tablename\ \thetable{} -- continued from previous page}} \\
\hline
Package & Metric & MRR & 2 & 3 & 4 & 5 & 6 & 7 & 8 \\
\hline
\endhead

\hline \multicolumn{10}{r}{{Continued on next page}} \\
\endfoot

\hline
\endlastfoot
Spec2-Adapters-Morphic & without & 0.38 & 0.04 & 0.18 & 0.31 & 0.43 & 0.59 & 0.62 & 0.66 \\
Spec2-Adapters-Morphic & with & 0.45 & 0.05 & 0.29 & 0.39 & 0.5 & 0.65 & 0.69 & 0.72 \\
\hline
Spec2-Adapters-Morphic-ListView & without & 0.31 & 0.0 & 0.05 & 0.21 & 0.46 & 0.49 & 0.49 & 0.49 \\
Spec2-Adapters-Morphic-ListView & with & 0.40& 0.09 & 0.15 & 0.28 & 0.58 & 0.58 & 0.58 & 0.58 \\
\hline
Spec2-Adapters-Morphic-Tests & without & 0.29 & 0.0 & 0.06 & 0.2 & 0.36 & 0.4 & 0.41 & 0.81 \\
Spec2-Adapters-Morphic-Tests & with & 0.22 & 0.0 & 0.04 & 0.15 & 0.28 & 0.31 & 0.33 & 0.55 \\
\hline
Spec2-Adapters-Stub & without & 0.16 & 0.0 & 0.02 & 0.02 & 0.02 & 0.02 & 0.41 & 0.64 \\
Spec2-Adapters-Stub & with & 0.15 & 0.03 & 0.03 & 0.03 & 0.03 & 0.03 & 0.35 & 0.55 \\
\hline
Spec2-Backend-Tests & without & 0.26& 0.0 & 0.04 & 0.22 & 0.3 & 0.36 & 0.44 & 0.52 \\
Spec2-Backend-Tests & with & 0.34 & 0.0 & 0.12 & 0.36 & 0.39 & 0.46 & 0.53 & 0.58 \\
\hline
Spec2-Code & without & 0.26 & 0.09 & 0.14 & 0.2 & 0.24 & 0.24 & 0.49 & 0.55 \\
Spec2-Code & with & 0.30 & 0.1 & 0.17 & 0.22 & 0.25 & 0.28 & 0.57 & 0.6 \\
\hline
Spec2-Code-Backend-Tests & without & 0.15 & 0.02 & 0.02 & 0.02 & 0.11 & 0.11 & 0.37 & 0.64 \\
Spec2-Code-Backend-Tests & with & 0.35 & 0.02 & 0.21 & 0.21 & 0.34 & 0.34 & 0.72 & 0.83 \\
\hline
Spec2-Code-Commands & without & 0.12 & 0.0 & 0.02 & 0.06 & 0.16 & 0.17 & 0.21 & 0.31 \\
Spec2-Code-Commands & with & 0.09 & 0.0 & 0.02 & 0.02 & 0.12 & 0.13 & 0.18 & 0.21 \\
\hline
Spec2-Code-Diff & without & 0.45& 0.14 & 0.24 & 0.25 & 0.57 & 0.57 & 0.81 & 0.88 \\
Spec2-Code-Diff & with & 0.47 & 0.21 & 0.3 & 0.31 & 0.55 & 0.55 & 0.77 & 0.86 \\
\hline
Spec2-Code-Diff-Morphic & without & 0.31 & 0.02 & 0.06 & 0.18 & 0.23 & 0.53 & 0.65 & 0.68 \\
Spec2-Code-Diff-Morphic & with & 0.48 & 0.17 & 0.2 & 0.36 & 0.45 & 0.67 & 0.82 & 0.86 \\
\hline
Spec2-Code-Diff-Tests & without & 0.62 & 0.07 & 0.26 & 0.26 & 1.0 & 1.0 & 1.0 & 1.0 \\
Spec2-Code-Diff-Tests & with & 0.54 & 0.15 & 0.29 & 0.29 & 0.84 & 0.8 & 0.8 & 0.8 \\
\hline
Spec2-Code-Morphic & without & 0.28 & 0.0 & 0.27 & 0.3 & 0.31 & 0.36 & 0.36 & 0.36 \\
Spec2-Code-Morphic & with & 0.45& 0.37 & 0.42 & 0.45 & 0.45 & 0.51 & 0.51 & 0.51 \\
\hline
Spec2-Code-Tests & without & 0.27 & 0.03 & 0.06 & 0.14 & 0.21 & 0.21 & 0.59 & 0.78 \\
Spec2-Code-Tests & with & 0.37 & 0.1 & 0.18 & 0.22 & 0.35 & 0.35 & 0.71 & 0.8 \\
\hline
Spec2-CommandLine & without & 0.18 & 0.03 & 0.03 & 0.2 & 0.2 & 0.28 & 0.28 & 0.28 \\
Spec2-CommandLine & with & 0.30 & 0.03 & 0.03 & 0.37 & 0.37 & 0.44 & 0.44 & 0.44 \\
\hline
Spec2-Commander2 & without & 0.33 & 0.03 & 0.15 & 0.41 & 0.43 & 0.42 & 0.5 & 0.5 \\
Spec2-Commander2 & with & 0.46 & 0.14 & 0.44 & 0.5 & 0.52 & 0.52 & 0.58 & 0.58 \\
\hline
Spec2-Commander2-Tests & without & 0.60 & 0.01 & 0.64 & 0.68 & 0.71 & 0.72 & 0.75 & 0.76 \\
Spec2-Commander2-Tests & with & 0.59 & 0.0 & 0.64 & 0.67 & 0.69 & 0.69 & 0.75 & 0.76 \\
\hline
Spec2-Commands & without & 0.38 & 0.24 & 0.24 & 0.28 & 0.31 & 0.35 & 0.49 & 0.88 \\
Spec2-Commands & with & 0.51 & 0.24 & 0.32 & 0.51 & 0.53 & 0.62 & 0.62 & 0.88 \\
\hline
Spec2-CommonWidgets & without & 0.26 & 0.03 & 0.11 & 0.28 & 0.34 & 0.34 & 0.35 & 0.43 \\
Spec2-CommonWidgets & with & 0.37 & 0.04 & 0.22 & 0.36 & 0.4 & 0.49 & 0.51 & 0.61 \\
\hline
Spec2-Core & without & 0.37 & 0.08 & 0.17 & 0.35 & 0.43 & 0.47 & 0.52 & 0.63 \\
Spec2-Core & with & 0.495 & 0.08 & 0.3 & 0.5 & 0.58 & 0.63 & 0.65 & 0.76 \\
\hline
Spec2-Dialogs & without & 0.18 & 0.02 & 0.03 & 0.21 & 0.24 & 0.24 & 0.25 & 0.29 \\
Spec2-Dialogs & with & 0.33 & 0.02 & 0.18 & 0.33 & 0.35 & 0.48 & 0.48 & 0.53 \\
\hline
Spec2-Dialogs-Tests & without & 0.28& 0.05 & 0.13 & 0.19 & 0.34 & 0.56 & 0.56 & 0.56 \\
Spec2-Dialogs-Tests & with & 0.27 & 0.05 & 0.12 & 0.19 & 0.31 & 0.56 & 0.56 & 0.56 \\
\hline
Spec2-Examples & without & 0.18 & 0.0 & 0.02 & 0.19 & 0.24 & 0.24 & 0.28 & 0.33 \\
Spec2-Examples & with & 0.23 & 0.01 & 0.13 & 0.23 & 0.27 & 0.3 & 0.34 & 0.37 \\
\hline
Spec2-Interactions & without & 0 & 0.0 & 0.0 & 0.0 & 0.0 & 0.0 & 0.0 & 0.0 \\
Spec2-Interactions & with & 0 & 0.0 & 0.0 & 0.0 & 0.0 & 0.0 & 0.0 & 0.0 \\
\hline
Spec2-Layout & without & 0.49 & 0.03 & 0.14 & 0.38 & 0.63 & 0.72 & 0.77 & 0.83 \\
Spec2-Layout & with & 0.42 & 0.04 & 0.42 & 0.45 & 0.45 & 0.49 & 0.53 & 0.58 \\
\hline
Spec2-ListView & without & 0.27 & 0.02 & 0.07 & 0.29 & 0.31 & 0.34 & 0.44 & 0.44 \\
Spec2-ListView & with & 0.37 & 0.03 & 0.18 & 0.38 & 0.39 & 0.46 & 0.57 & 0.57 \\
\hline
Spec2-ListView-Tests & without & 0.25 & 0.03 & 0.05 & 0.08 & 0.25 & 0.34 & 0.52 & 0.52 \\
Spec2-ListView-Tests & with & 0.24 & 0.03 & 0.03 & 0.21 & 0.24 & 0.32 & 0.44 & 0.44 \\
\hline
Spec2-Microdown & without & 0 & 0.0 & 0.0 & 0.0 & 0.0 & 0.0 & 0.0 & 0.0 \\
Spec2-Microdown & with & 0 & 0.0 & 0.0 & 0.0 & 0.0 & 0.0 & 0.0 & 0.0 \\
\hline
Spec2-Morphic & without & 0.40 & 0.13 & 0.3 & 0.37 & 0.44 & 0.49 & 0.54 & 0.73 \\
Spec2-Morphic & with & 0.47 & 0.18 & 0.35 & 0.46 & 0.55 & 0.6 & 0.64 & 0.76 \\
\hline
Spec2-Morphic-Backend-Tests & without & 0.19 & 0.01 & 0.05 & 0.14 & 0.21 & 0.25 & 0.32 & 0.41 \\
Spec2-Morphic-Backend-Tests & with & 0.29 & 0.04 & 0.15 & 0.26 & 0.34 & 0.4 & 0.44 & 0.51 \\
\hline
Spec2-Morphic-Examples & without & 0.11 & 0.0 & 0.03 & 0.11 & 0.13 & 0.21 & 0.24 & 0.24 \\
Spec2-Morphic-Examples & with & 0.17 & 0.0 & 0.08 & 0.15 & 0.17 & 0.35 & 0.37 & 0.37 \\
\hline
Spec2-Morphic-Tests & without & 0.38 & 0.05 & 0.21 & 0.34 & 0.42 & 0.53 & 0.71 & 0.74 \\
Spec2-Morphic-Tests & with & 0.39 & 0.1 & 0.28 & 0.4 & 0.45 & 0.57 & 0.58 & 0.6 \\
\hline
Spec2-Tests & without & 0.23 & 0.01 & 0.06 & 0.19 & 0.28 & 0.29 & 0.36 & 0.45 \\
Spec2-Tests & with & 0.21 & 0.01 & 0.07 & 0.2 & 0.26 & 0.28 & 0.34 & 0.38 \\
\hline
Spec2-Transmission & without & 0.183 & 0.02 & 0.02 & 0.12 & 0.25 & 0.26 & 0.26 & 0.36 \\
Spec2-Transmission & with & 0.75 & 0.07 & 0.33 & 0.98 & 0.98 & 0.98 & 0.98 & 0.98 \\
\end{longtable}

\newpage

\section{Roassal}
\scriptsize
\begin{longtable}{llllllllll}
  \caption{Performance of Roassal Framework}\\
\hline
Package & Metric & MRR & 2 & 3 & 4 & 5 & 6 & 7 & 8 \\
\hline
\endfirsthead

\multicolumn{10}{c}%
{{\bfseries \tablename\ \thetable{} -- continued from previous page}} \\
\hline
Package & Metric & MRR & 2 & 3 & 4 & 5 & 6 & 7 & 8 \\
\hline
\endhead

\hline \multicolumn{10}{r}{{Continued on next page}} \\
\endfoot

\hline
\endlastfoot

Roassal & without & 0.29 & 0.05 & 0.13 & 0.22 & 0.32 & 0.4 & 0.44 & 0.64 \\
Roassal & with & 0.31 & 0.08 & 0.18 & 0.27 & 0.32 & 0.39 & 0.42 & 0.64 \\
\hline
Roassal-Animation & without & 0.41 & 0.09 & 0.38 & 0.43 & 0.45 & 0.57 & 0.59 & 0.53 \\
Roassal-Animation & with & 0.54 & 0.15 & 0.55 & 0.55 & 0.57 & 0.74 & 0.74 & 0.73 \\
\hline
Roassal-Animation-Tests & without & 0.33 & 0.01 & 0.05 & 0.34 & 0.46 & 0.47 & 0.47 & 0.57 \\
Roassal-Animation-Tests & with & 0.27 & 0.01 & 0.03 & 0.21 & 0.39 & 0.41 & 0.41 & 0.47 \\
\hline
Roassal-BaselineMap & without & 0.31 & 0.02 & 0.08 & 0.19 & 0.36 & 0.42 & 0.59 & 0.64 \\
Roassal-BaselineMap & with & 0.33 & 0.06 & 0.14 & 0.27 & 0.41 & 0.45 & 0.5 & 0.55 \\
\hline
Roassal-BaselineMap-Tests & without & 0.30 & 0.0 & 0.07 & 0.36 & 0.39 & 0.39 & 0.44 & 0.5 \\
Roassal-BaselineMap-Tests & with & 0.25& 0.0 & 0.13 & 0.33 & 0.33 & 0.33 & 0.33 & 0.33 \\
\hline
Roassal-Builders & without & 0.40 & 0.04 & 0.19 & 0.38 & 0.57 & 0.61 & 0.66 & 0.73 \\
Roassal-Builders & with & 0.29 & 0.04 & 0.15 & 0.3 & 0.43 & 0.42 & 0.45 & 0.51 \\
\hline
Roassal-Chart & without & 0.49 & 0.04 & 0.31 & 0.49 & 0.61 & 0.67 & 0.76 & 0.81 \\
Roassal-Chart & with & 0.47 & 0.05 & 0.31 & 0.45 & 0.59 & 0.65 & 0.69 & 0.78 \\
\hline
Roassal-Chart-Examples & without & 0.33 & 0.03 & 0.13 & 0.29 & 0.44 & 0.48 & 0.55 & 0.61 \\
Roassal-Chart-Examples & with & 0.29 & 0.03 & 0.13 & 0.26 & 0.39 & 0.42 & 0.47 & 0.51 \\
\hline
Roassal-Chart-Tests & without & 0.48 & 0.04 & 0.25 & 0.48 & 0.57 & 0.67 & 0.76 & 0.81 \\
Roassal-Chart-Tests & with & 0.30 & 0.04 & 0.12 & 0.26 & 0.38 & 0.43 & 0.48 & 0.5 \\
\hline
Roassal-Class-Examples & without & 0 & 0.0 & 0.0 & 0.0 & 0.0 & 0.0 & 0.0 & 0.0 \\
Roassal-Class-Examples & with & 0 & 0.0 & 0.0 & 0.0 & 0.0 & 0.0 & 0.0 & 0.0 \\
\hline
Roassal-Colors & without & 0.52 & 0.11 & 0.51 & 0.55 & 0.55 & 0.78 & 0.78 & 1.0 \\
Roassal-Colors & with & 0.52 & 0.11 & 0.51 & 0.55 & 0.55 & 0.78 & 0.78 & 1.0 \\
\hline
Roassal-DSM & without & 0.48 & 0.01 & 0.14 & 0.43 & 0.66 & 0.88 & 0.88 & 0.85 \\
Roassal-DSM & with & 0.32 & 0.01 & 0.09 & 0.34 & 0.48 & 0.51 & 0.51 & 0.53 \\
\hline
Roassal-Event & without & 0.21 & 0.13 & 0.16 & 0.22 & 0.22 & 0.27 & 0.27 & 0.27 \\
Roassal-Event & with & 0.21 & 0.13 & 0.16 & 0.22 & 0.22 & 0.27 & 0.27 & 0.27 \\
\hline
Roassal-Examples & without & 0.34 & 0.01 & 0.11 & 0.25 & 0.44 & 0.53 & 0.65 & 0.71 \\
Roassal-Examples & with & 0.36 & 0.01 & 0.11 & 0.29 & 0.5 & 0.57 & 0.63 & 0.7 \\
\hline
Roassal-Experimental & without & 0.37 & 0.06 & 0.26 & 0.34 & 0.41 & 0.48 & 0.51 & 0.62 \\
Roassal-Experimental & with & 0.37 & 0.07 & 0.27 & 0.35 & 0.43 & 0.49 & 0.51 & 0.62 \\
\hline
Roassal-Exporters & without & 0.29 & 0.01 & 0.3 & 0.33 & 0.34 & 0.37 & 0.37 & 0.37 \\
Roassal-Exporters & with & 0.31 & 0.02 & 0.33 & 0.34 & 0.36 & 0.38 & 0.38 & 0.39 \\
\hline
Roassal-Exporters-Examples & without & 0.36 & 0.01 & 0.08 & 0.22 & 0.53 & 0.6 & 0.6 & 0.71 \\
Roassal-Exporters-Examples & with & 0.38 & 0.01 & 0.09 & 0.26 & 0.54 & 0.61 & 0.61 & 0.75 \\
\hline
Roassal-Exporters-Tests & without & 0.28 & 0.01 & 0.05 & 0.19 & 0.4 & 0.46 & 0.55 & 0.64 \\
Roassal-Exporters-Tests & with & 0.30 & 0.01 & 0.06 & 0.25 & 0.45 & 0.49 & 0.57 & 0.66 \\
\hline
Roassal-FlameGraph-Examples & without & 0.41 & 0.0 & 0.1 & 0.38 & 0.58 & 0.62 & 0.67 & 0.68 \\
Roassal-FlameGraph-Examples & with & 0.29 & 0.0 & 0.06 & 0.32 & 0.4 & 0.42 & 0.46 & 0.48 \\
\hline
Roassal-Global-Tests & without & 0.28 & 0.02 & 0.09 & 0.24 & 0.36 & 0.48 & 0.55 & 0.66 \\
Roassal-Global-Tests & with & 0.28 & 0.01 & 0.07 & 0.21 & 0.43 & 0.49 & 0.52 & 0.68 \\
\hline
Roassal-Inspector & without & 0.21 & 0.0 & 0.05 & 0.2 & 0.28 & 0.29 & 0.32 & 0.39 \\
Roassal-Inspector & with & 0.19 & 0.04 & 0.07 & 0.16 & 0.25 & 0.25 & 0.27 & 0.37 \\
\hline
Roassal-Inspector-Tests & without & 0.25 & 0.0 & 0.08 & 0.2 & 0.35 & 0.43 & 0.43 & 0.5 \\
Roassal-Inspector-Tests & with & 0.31 & 0.0 & 0.11 & 0.22 & 0.49 & 0.49 & 0.49 & 0.64 \\
\hline
Roassal-Interaction & without & 0.35 & 0.01 & 0.13 & 0.3 & 0.44 & 0.53 & 0.57 & 0.7 \\
Roassal-Interaction & with & 0.35 & 0.02 & 0.11 & 0.28 & 0.47 & 0.52 & 0.55 & 0.71 \\
\hline
Roassal-Interaction-Tests & without & 0.26 & 0.0 & 0.06 & 0.18 & 0.33 & 0.4 & 0.48 & 0.59 \\
Roassal-Interaction-Tests & with & 0.28 & 0.0 & 0.08 & 0.22 & 0.38 & 0.42 & 0.44 & 0.58 \\
\hline
Roassal-LayoutStudio & without & 0.24 & 0.01 & 0.08 & 0.25 & 0.32 & 0.32 & 0.33 & 0.4 \\
Roassal-LayoutStudio & with & 0.26 & 0.05 & 0.17 & 0.29 & 0.31 & 0.32 & 0.33 & 0.4 \\
\hline
Roassal-LayoutStudio-Tests & without & 0.49 & 0.0 & 0.19 & 0.25 & 0.75 & 0.75 & 0.75 & 0.75 \\
Roassal-LayoutStudio-Tests & with & 0.27 & 0.0 & 0.19 & 0.25 & 0.38 & 0.38 & 0.38 & 0.38 \\
\hline
Roassal-Layouts & without & 0.42 & 0.11 & 0.2 & 0.35 & 0.49 & 0.57 & 0.61 & 0.78 \\
Roassal-Layouts & with & 0.52 & 0.12 & 0.37 & 0.51 & 0.59 & 0.66 & 0.7 & 0.87 \\
\hline
Roassal-Layouts-Tests & without & 0.34 & 0.01 & 0.13 & 0.3 & 0.47 & 0.48 & 0.54 & 0.54 \\
Roassal-Layouts-Tests & with & 0.20 & 0.01 & 0.08 & 0.24 & 0.28 & 0.26 & 0.29 & 0.29 \\
\hline
Roassal-Layouts-Util & without & 0.43 & 0.22 & 0.41 & 0.49 & 0.51 & 0.42 & 0.42 & 0.62 \\
Roassal-Layouts-Util & with & 0.44& 0.22 & 0.44 & 0.5 & 0.52 & 0.44 & 0.44 & 0.63 \\
\hline
Roassal-Legend & without & 0.27 & 0.0 & 0.09 & 0.23 & 0.38 & 0.44 & 0.5 & 0.64 \\
Roassal-Legend & with & 0.22 & 0.0 & 0.05 & 0.19 & 0.35 & 0.36 & 0.38 & 0.55 \\
\hline
Roassal-Legend-Examples & without & 0.35 & 0.0 & 0.1 & 0.28 & 0.44 & 0.49 & 0.69 & 0.71 \\
Roassal-Legend-Examples & with & 0.28& 0.0 & 0.1 & 0.28 & 0.4 & 0.41 & 0.47 & 0.47 \\
\hline
Roassal-Menu & without & 0 & 0.0 & 0.0 & 0.0 & 0.0 & 0.0 & 0.0 & 0.0 \\
Roassal-Menu & with & 0 & 0.0 & 0.0 & 0.0 & 0.0 & 0.0 & 0.0 & 0.0 \\
\hline
Roassal-Mondrian & without & 0.36 & 0.01 & 0.15 & 0.28 & 0.46 & 0.55 & 0.6 & 0.66 \\
Roassal-Mondrian & with & 0.33 & 0.04 & 0.19 & 0.3 & 0.43 & 0.46 & 0.48 & 0.55 \\
\hline
Roassal-Pie & without & 0.49 & 0.03 & 0.26 & 0.51 & 0.58 & 0.74 & 0.74 & 0.93 \\
Roassal-Pie & with & 0.36 & 0.03 & 0.14 & 0.29 & 0.38 & 0.56 & 0.63 & 0.86 \\
\hline
Roassal-Pie-Examples & without & 0.36 & 0.03 & 0.19 & 0.32 & 0.45 & 0.58 & 0.62 & 0.69 \\
Roassal-Pie-Examples & with & 0.35 & 0.03 & 0.17 & 0.32 & 0.46 & 0.57 & 0.59 & 0.68 \\
\hline
Roassal-SVG & without & 0.40 & 0.13 & 0.24 & 0.32 & 0.35 & 0.4 & 0.6 & 0.87 \\
Roassal-SVG & with & 0.43 & 0.17 & 0.3 & 0.37 & 0.4 & 0.4 & 0.6 & 0.87 \\
\hline
Roassal-SVG-Examples & without & 0.42 & 0.02 & 0.12 & 0.26 & 0.45 & 0.71 & 0.77 & 0.86 \\
Roassal-SVG-Examples & with & 0.42& 0.01 & 0.09 & 0.27 & 0.46 & 0.73 & 0.76 & 0.86 \\
\hline
Roassal-SVG-Tests & without & 0.33 & 0.0 & 0.02 & 0.23 & 0.37 & 0.7 & 0.73 & 0.73 \\
Roassal-SVG-Tests & with & 0.36 & 0.0 & 0.03 & 0.24 & 0.46 & 0.74 & 0.76 & 0.76 \\
\hline
Roassal-Shapes & without & 0.49 & 0.08 & 0.23 & 0.42 & 0.62 & 0.71 & 0.77 & 0.84 \\
Roassal-Shapes & with & 0.54 & 0.12 & 0.42 & 0.51 & 0.65 & 0.74 & 0.77 & 0.84 \\
\hline
Roassal-Shapes-Tests & without & 0.31 & 0.03 & 0.12 & 0.26 & 0.4 & 0.44 & 0.71 & 0.74 \\
Roassal-Shapes-Tests & with & 0.33 & 0.03 & 0.11 & 0.29 & 0.48 & 0.5 & 0.61 & 0.68 \\
\hline
Roassal-Spec & without & 0.45 & 0.0 & 0.25 & 0.5 & 0.6 & 0.6 & 0.6 & 0.6 \\
Roassal-Spec & with & 0.85  & 0.55 & 0.67 & 0.75 & 1.0 & 1.0 & 1.0 & 1.0 \\
\hline
Roassal-Spec-Examples & without & 0.39 & 0.01 & 0.16 & 0.32 & 0.51 & 0.55 & 0.63 & 0.81 \\
Roassal-Spec-Examples & with & 0.34 & 0.01 & 0.09 & 0.27 & 0.48 & 0.5 & 0.55 & 0.71 \\
\hline
Roassal-Spec-Morphic & without & 0.11 & 0.0 & 0.0 & 0.0 & 0.2 & 0.2 & 0.2 & 0.2 \\
Roassal-Spec-Morphic & with & 0.70 & 0.1 & 0.33 & 0.5 & 1.0 & 1.0 & 1.0 & 1.0 \\
\hline
Roassal-Spec-Tests & without & 0.31 & 0.01 & 0.12 & 0.18 & 0.32 & 0.56 & 0.56 & 0.56 \\
Roassal-Spec-Tests & with & 0.42 & 0.06 & 0.22 & 0.3 & 0.47 & 0.69 & 0.69 & 0.69 \\
\hline
Roassal-Sunburst & without & 0.32 & 0.06 & 0.21 & 0.35 & 0.36 & 0.4 & 0.4 & 0.51 \\
Roassal-Sunburst & with & 0.48 & 0.27 & 0.41 & 0.46 & 0.46 & 0.57 & 0.57 & 0.71 \\
\hline
Roassal-Sunburst-Examples & without & 0.36 & 0.0 & 0.08 & 0.24 & 0.47 & 0.52 & 0.68 & 0.77 \\
Roassal-Sunburst-Examples & with & 0.281 & 0.0 & 0.08 & 0.23 & 0.37 & 0.42 & 0.47 & 0.52 \\
\hline
Roassal-TreeMap & without & 0.35 & 0.08 & 0.23 & 0.34 & 0.43 & 0.46 & 0.52 & 0.61 \\
Roassal-TreeMap & with & 0.50 & 0.22 & 0.42 & 0.46 & 0.56 & 0.61 & 0.66 & 0.76 \\
\hline
Roassal-TreeMap-Examples & without & 0.28 & 0.01 & 0.1 & 0.22 & 0.4 & 0.46 & 0.5 & 0.57 \\
Roassal-TreeMap-Examples & with & 0.27 & 0.01 & 0.09 & 0.23 & 0.38 & 0.44 & 0.46 & 0.52 \\
\hline
Roassal-UML & without & 0.24& 0.0 & 0.06 & 0.16 & 0.34 & 0.39 & 0.45 & 0.59 \\
Roassal-UML & with & 0.28 & 0.01 & 0.08 & 0.2 & 0.44 & 0.45 & 0.47 & 0.6 \\
\hline
Roassal-UML-Calypso & without & 0.35 & 0.03 & 0.15 & 0.27 & 0.43 & 0.51 & 0.59 & 0.72 \\
Roassal-UML-Calypso & with & 0.31& 0.03 & 0.13 & 0.25 & 0.4 & 0.48 & 0.51 & 0.62 \\
\hline
Roassal-UML-Examples & without & 0.31 & 0.0 & 0.07 & 0.22 & 0.33 & 0.37 & 0.6 & 0.75 \\
Roassal-UML-Examples & with & 0.26 & 0.01 & 0.04 & 0.15 & 0.32 & 0.38 & 0.45 & 0.58 \\
\hline
Roassal-UML-Tests & without & 0.25 & 0.0 & 0.03 & 0.21 & 0.35 & 0.4 & 0.43 & 0.57 \\
Roassal-UML-Tests & with & 0.30 & 0.0 & 0.0 & 0.18 & 0.48 & 0.49 & 0.51 & 0.75 \\
\hline

\end{longtable}

\newpage 
\section{IceBerg}
\scriptsize
\begin{longtable}{llllllllll}
  \caption{Performance of Iceberg Library }\\
\hline
Package & Metric & MRR & 2 & 3 & 4 & 5 & 6 & 7 & 8 \\
\hline
\endfirsthead

\multicolumn{10}{c}%
{{\bfseries \tablename\ \thetable{} -- continued from previous page}} \\
\hline
Package & Metric & MRR & 2 & 3 & 4 & 5 & 6 & 7 & 8 \\
\hline
\endhead

\hline \multicolumn{10}{r}{{Continued on next page}} \\
\endfoot

\hline
\endlastfoot

Iceberg & without & 0.38 & 0.03 & 0.08 & 0.23 & 0.43 & 0.6 & 0.67 & 0.69 \\
Iceberg & with & 0.39 & 0.04 & 0.08 & 0.27 & 0.45 & 0.61 & 0.67 & 0.73 \\
\hline
Iceberg-Libgit & without & 0.30 & 0.02 & 0.05 & 0.15 & 0.32 & 0.43 & 0.55 & 0.64 \\
Iceberg-Libgit & with & 0.32 & 0.04 & 0.06 & 0.21 & 0.34 & 0.45 & 0.57 & 0.66 \\
\hline
Iceberg-Libgit-Filetree & without & 0.26 & 0.08 & 0.1 & 0.21 & 0.3 & 0.33 & 0.34 & 0.5 \\
Iceberg-Libgit-Filetree & with & 0.40 & 0.29 & 0.31 & 0.38 & 0.41 & 0.41 & 0.43 & 0.59 \\
\hline
Iceberg-Libgit-Tonel & without & 0.16 & 0.04 & 0.06 & 0.1 & 0.2 & 0.22 & 0.23 & 0.31 \\
Iceberg-Libgit-Tonel & with & 0.32 & 0.21 & 0.24 & 0.28 & 0.35 & 0.37 & 0.39 & 0.47 \\
\hline
Iceberg-Metacello-Integration & without & 0.26 & 0.0 & 0.02 & 0.3 & 0.31 & 0.42 & 0.44 & 0.39 \\
Iceberg-Metacello-Integration & with & 0.45 & 0.15 & 0.16 & 0.57 & 0.58 & 0.6 & 0.62 & 0.56 \\
\hline
Iceberg-Plugin & without & 0.57 & 0.22 & 0.67 & 0.71 & 0.78 & 0.5 & 0.5 & 0.5 \\
Iceberg-Plugin & with & 0.63 & 0.29 & 0.73 & 0.83 & 0.83 & 0.5 & 0.5 & 0.5 \\
\hline
Iceberg-Plugin-GitHub & without & 0.26 & 0.02 & 0.08 & 0.18 & 0.36 & 0.36 & 0.43 & 0.45 \\
Iceberg-Plugin-GitHub & with & 0.28 & 0.06 & 0.12 & 0.21 & 0.38 & 0.38 & 0.45 & 0.47 \\
\hline
Iceberg-Plugin-Migration & without & 0.27 & 0.0 & 0.05 & 0.2 & 0.28 & 0.43 & 0.43 & 0.51 \\
Iceberg-Plugin-Migration & with & 0.32 & 0.13 & 0.17 & 0.35 & 0.36 & 0.36 & 0.43 & 0.47 \\
\hline
Iceberg-Plugin-Pharo & without & 0.51 & 0.1 & 0.5 & 0.5 & 0.5 & 0.5 & 0.75 & 0.75 \\
Iceberg-Plugin-Pharo & with & 0.51 & 0.1 & 0.5 & 0.5 & 0.5 & 0.5 & 0.75 & 0.75 \\
\hline
Iceberg-TipUI & without & 0.21 & 0.01 & 0.04 & 0.15 & 0.28 & 0.29 & 0.34 & 0.41 \\
Iceberg-TipUI & with & 0.23 & 0.01 & 0.04 & 0.16 & 0.28 & 0.29 & 0.39 & 0.47 \\
\hline
Iceberg-TipUI-SnapshotBrowser & without & 0.26 & 0.07 & 0.21 & 0.33 & 0.36 & 0.31 & 0.27 & 0.34 \\
Iceberg-TipUI-SnapshotBrowser & with & 0.26 & 0.07 & 0.21 & 0.33 & 0.36 & 0.31 & 0.27 & 0.34 \\
\hline

\end{longtable}

\newpage
\section{Moose}
\scriptsize
\begin{longtable}{llllllllll}
  \caption{Performance of Moose Framework }\\
\hline
Package & Metric & MRR & 2 & 3 & 4 & 5 & 6 & 7 & 8 \\
\hline
\endfirsthead

\multicolumn{10}{c}%
{{\bfseries \tablename\ \thetable{} -- continued from previous page}} \\
\hline
Package & Metric & MRR & 2 & 3 & 4 & 5 & 6 & 7 & 8 \\
\hline
\endhead

\hline \multicolumn{10}{r}{{Continued on next page}} \\
\endfoot

\hline
\endlastfoot

Moose-Blueprint-Invocations-Models & without & 0.428 & 0.02 & 0.49 & 0.49 & 0.49 & 0.49 & 0.49 & 0.53 \\
Moose-Blueprint-Invocations-Models & with & 0.846 & 0.83 & 0.84 & 0.84 & 0.84 & 0.84 & 0.84 & 0.89 \\
\hline
Moose-Blueprint-Models & without & 0.373 & 0.03 & 0.19 & 0.39 & 0.49 & 0.52 & 0.52 & 0.62 \\
Moose-Blueprint-Models & with & 0.397 & 0.03 & 0.23 & 0.42 & 0.51 & 0.55 & 0.55 & 0.66 \\
\hline
Moose-Blueprint-Models-Tests & without & 0.089 & 0.0 & 0.04 & 0.06 & 0.06 & 0.08 & 0.08 & 0.31 \\
Moose-Blueprint-Models-Tests & with & 0.086 & 0.0 & 0.06 & 0.06 & 0.06 & 0.06 & 0.06 & 0.31 \\
\hline
Moose-Blueprint-Visualization-Models & without & 0.407 & 0.06 & 0.2 & 0.43 & 0.51 & 0.58 & 0.58 & 0.65 \\
Moose-Blueprint-Visualization-Models & with & 0.410 & 0.06 & 0.21 & 0.43 & 0.51 & 0.58 & 0.6 & 0.67 \\
\hline
Moose-Configuration & without & 0 & 0.0 & 0.0 & 0.0 & 0.0 & 0.0 & 0.0 & 0.0 \\
Moose-Configuration & with & 0 & 0.0 & 0.0 & 0.0 & 0.0 & 0.0 & 0.0 & 0.0 \\
\hline
Moose-Core & without & 0.338 & 0.12 & 0.2 & 0.24 & 0.27 & 0.45 & 0.59 & 0.63 \\
Moose-Core & with & 0.396 & 0.16 & 0.24 & 0.29 & 0.32 & 0.57 & 0.63 & 0.67 \\
\hline
Moose-Core-Generator & without & 0.158 & 0.13 & 0.17 & 0.17 & 0.17 & 0.17 & 0.0 & 0.0 \\
Moose-Core-Generator & with & 0.158 & 0.13 & 0.17 & 0.17 & 0.17 & 0.17 & 0.0 & 0.0 \\
\hline
Moose-Core-Tests & without & 0.146 & 0.03 & 0.05 & 0.07 & 0.07 & 0.24 & 0.29 & 0.32 \\
Moose-Core-Tests & with & 0.137 & 0.03 & 0.05 & 0.06 & 0.06 & 0.22 & 0.28 & 0.3 \\
\hline
Moose-Core-Tests-Entities & without & 0 & 0.0 & 0.0 & 0.0 & 0.0 & 0.0 & 0.0 & 0.0 \\
Moose-Core-Tests-Entities & with & 0 & 0.0 & 0.0 & 0.0 & 0.0 & 0.0 & 0.0 & 0.0 \\
\hline
Moose-Importers & without & 0.198 & 0.02 & 0.05 & 0.08 & 0.04 & 0.11 & 0.44 & 0.71 \\
Moose-Importers & with & 0.280 & 0.14 & 0.18 & 0.21 & 0.19 & 0.19 & 0.42 & 0.68 \\
\hline
Moose-Importers-Tests & without & 0.080 & 0.03 & 0.03 & 0.06 & 0.08 & 0.11 & 0.11 & 0.19 \\
Moose-Importers-Tests & with & 0.117 & 0.07 & 0.07 & 0.1 & 0.12 & 0.14 & 0.14 & 0.22 \\
\hline
Moose-Query & without & 0.293 & 0.1 & 0.25 & 0.25 & 0.25 & 0.27 & 0.27 & 0.67 \\
Moose-Query & with & 0.332 & 0.15 & 0.3 & 0.3 & 0.3 & 0.31 & 0.31 & 0.67 \\
\hline
Moose-Query-Extensions & without & 0 & 0.0 & 0.0 & 0.0 & 0.0 & 0.0 & 0.0 & 0.0 \\
Moose-Query-Extensions & with & 0 & 0.0 & 0.0 & 0.0 & 0.0 & 0.0 & 0.0 & 0.0 \\
\hline
Moose-Query-Test & without & 0.105 & 0.0 & 0.01 & 0.01 & 0.01 & 0.04 & 0.25 & 0.43 \\
Moose-Query-Test & with & 0.092 & 0.0 & 0.0 & 0.0 & 0.0 & 0.03 & 0.21 & 0.4 \\
\hline
Moose-SmalltalkImporter & without & 0.262 & 0.12 & 0.19 & 0.22 & 0.27 & 0.3 & 0.34 & 0.45 \\
Moose-SmalltalkImporter & with & 0.345 & 0.22 & 0.29 & 0.32 & 0.37 & 0.36 & 0.41 & 0.49 \\
\hline
Moose-SmalltalkImporter-Core-Tests & without & 0.326 & 0.04 & 0.11 & 0.33 & 0.33 & 0.38 & 0.5 & 0.72 \\
Moose-SmalltalkImporter-Core-Tests & with & 0.487 & 0.37 & 0.38 & 0.39 & 0.39 & 0.49 & 0.61 & 0.88 \\
\hline
Moose-SmalltalkImporter-KGB-Tests & without & 0.175 & 0.02 & 0.02 & 0.03 & 0.03 & 0.03 & 0.46 & 0.66 \\
Moose-SmalltalkImporter-KGB-Tests & with & 0.178 & 0.02 & 0.02 & 0.03 & 0.03 & 0.03 & 0.46 & 0.66 \\
\hline
Moose-SmalltalkImporter-LAN-Tests & without & 0.359 & 0.1 & 0.1 & 0.39 & 0.41 & 0.44 & 0.51 & 0.67 \\
Moose-SmalltalkImporter-LAN-Tests & with & 0.380 & 0.03 & 0.03 & 0.39 & 0.42 & 0.51 & 0.57 & 0.88 \\
\hline
Moose-TestResources-KGB-P10InteractedReferee & without & 0 & 0.0 & 0.0 & 0.0 & 0.0 & 0.0 & 0.0 & 0.0 \\
Moose-TestResources-KGB-P10InteractedReferee & with & 0 & 0.0 & 0.0 & 0.0 & 0.0 & 0.0 & 0.0 & 0.0 \\
\hline
Moose-TestResources-KGB-P11FullReferee & without & 0 & 0.0 & 0.0 & 0.0 & 0.0 & 0.0 & 0.0 & 0.0 \\
Moose-TestResources-KGB-P11FullReferee & with & 0 & 0.0 & 0.0 & 0.0 & 0.0 & 0.0 & 0.0 & 0.0 \\
\hline
Moose-TestResources-KGB-P12FullReferencer & without & 0 & 0.0 & 0.0 & 0.0 & 0.0 & 0.0 & 0.0 & 0.0 \\
Moose-TestResources-KGB-P12FullReferencer & with & 0 & 0.0 & 0.0 & 0.0 & 0.0 & 0.0 & 0.0 & 0.0 \\
\hline
Moose-TestResources-KGB-P13FullReferencer & without & 0.785 & 0.5 & 0.5 & 0.5 & 1.0 & 1.0 & 1.0 & 1.0 \\
Moose-TestResources-KGB-P13FullReferencer & with & 0.785 & 0.5 & 0.5 & 0.5 & 1.0 & 1.0 & 1.0 & 1.0 \\
\hline
Moose-TestResources-KGB-P14FullReferee & without & 0 & 0.0 & 0.0 & 0.0 & 0.0 & 0.0 & 0.0 & 0.0 \\
Moose-TestResources-KGB-P14FullReferee & with & 0 & 0.0 & 0.0 & 0.0 & 0.0 & 0.0 & 0.0 & 0.0 \\
\hline
Moose-TestResources-KGB-P1FullReferencer & without & 0.666 & 0.33 & 0.33 & 0.67 & 0.67 & 0.67 & 1.0 & 1.0 \\
Moose-TestResources-KGB-P1FullReferencer & with & 0.704 & 0.47 & 0.47 & 0.67 & 0.67 & 0.67 & 1.0 & 1.0 \\
\hline
Moose-TestResources-KGB-P2InteractedReferencerReferee & without & 0.768 & 0.56 & 0.56 & 0.63 & 0.63 & 1.0 & 1.0 & 1.0 \\
Moose-TestResources-KGB-P2InteractedReferencerReferee & with & 0.942 & 0.9 & 0.9 & 0.9 & 0.9 & 1.0 & 1.0 & 1.0 \\
\hline
Moose-TestResources-KGB-P3InteractedReferencer & without & 0.800 & 0.49 & 0.49 & 0.81 & 0.81 & 1.0 & 1.0 & 1.0 \\
Moose-TestResources-KGB-P3InteractedReferencer & with & 0.6875 & 0.43 & 0.43 & 0.48 & 0.48 & 1.0 & 1.0 & 1.0 \\
\hline
Moose-TestResources-KGB-P4FullInteracted & without & 0.628 & 0.2 & 0.2 & 0.5 & 0.5 & 1.0 & 1.0 & 1.0 \\
Moose-TestResources-KGB-P4FullInteracted & with & 0.714 & 0.5 & 0.5 & 0.5 & 0.5 & 1.0 & 1.0 & 1.0 \\
\hline
Moose-TestResources-KGB-P5FullReferee & without & 0.857 & 0.5 & 0.5 & 1.0 & 1.0 & 1.0 & 1.0 & 1.0 \\
Moose-TestResources-KGB-P5FullReferee & with & 1.0 & 1.0 & 1.0 & 1.0 & 1.0 & 1.0 & 1.0 & 1.0 \\
\hline
Moose-TestResources-KGB-P6InteractedReferee & without & 0.571 & 0.25 & 0.25 & 0.5 & 0.5 & 0.5 & 1.0 & 1.0 \\
Moose-TestResources-KGB-P6InteractedReferee & with & 0.642 & 0.5 & 0.5 & 0.5 & 0.5 & 0.5 & 1.0 & 1.0 \\
\hline
Moose-TestResources-KGB-P7ReferencerReferee & without & 0.803 & 0.55 & 0.57 & 0.88 & 0.88 & 0.88 & 1.0 & 1.0 \\
Moose-TestResources-KGB-P7ReferencerReferee & with & 0.627 & 0.29 & 0.31 & 0.54 & 0.54 & 0.88 & 1.0 & 1.0 \\
\hline
Moose-TestResources-KGB-P8FullReferencer & without & 0.698 & 0.44 & 0.44 & 0.5 & 0.83 & 0.83 & 0.83 & 1.0 \\
Moose-TestResources-KGB-P8FullReferencer & with & 0.714 & 0.5 & 0.5 & 0.5 & 0.83 & 0.83 & 0.83 & 1.0 \\
\hline
Moose-TestResources-KGB-P9FullReferencer & without & 0.654 & 0.42 & 0.42 & 0.5 & 0.75 & 0.75 & 0.75 & 1.0 \\
Moose-TestResources-KGB-P9FullReferencer & with & 0.654 & 0.42 & 0.42 & 0.5 & 0.75 & 0.75 & 0.75 & 1.0 \\
\hline
Moose-TestResources-KGB-PExtensions & without & 0 & 0.0 & 0.0 & 0.0 & 0.0 & 0.0 & 0.0 & 0.0 \\
Moose-TestResources-KGB-PExtensions & with & 0 & 0.0 & 0.0 & 0.0 & 0.0 & 0.0 & 0.0 & 0.0 \\
\hline
Moose-TestResources-LAN & without & 0.337 & 0.08 & 0.08 & 0.31 & 0.38 & 0.51 & 0.53 & 0.5 \\
Moose-TestResources-LAN & with & 0.343 & 0.09 & 0.09 & 0.34 & 0.38 & 0.51 & 0.53 & 0.5 \\
\hline
Moose-TestResources-LCOM & without & 0 & 0.0 & 0.0 & 0.0 & 0.0 & 0.0 & 0.0 & 0.0 \\
Moose-TestResources-LCOM & with & 0 & 0.0 & 0.0 & 0.0 & 0.0 & 0.0 & 0.0 & 0.0 \\
\hline
Moose-TestResources-PackageBlueprint-P1 & without & 1.0 & 1.0 & 0.0 & 0.0 & 0.0 & 0.0 & 0.0 & 0.0 \\
Moose-TestResources-PackageBlueprint-P1 & with & 1.0 & 1.0 & 0.0 & 0.0 & 0.0 & 0.0 & 0.0 & 0.0 \\
\hline
Moose-TestResources-PackageBlueprint-P2 & without & 0 & 0.0 & 0.0 & 0.0 & 0.0 & 0.0 & 0.0 & 0.0 \\
Moose-TestResources-PackageBlueprint-P2 & with & 0 & 0.0 & 0.0 & 0.0 & 0.0 & 0.0 & 0.0 & 0.0 \\
\hline
Moose-TestResources-PackageBlueprint-P3 & without & 1.0 & 1.0 & 0.0 & 0.0 & 0.0 & 0.0 & 0.0 & 0.0 \\
Moose-TestResources-PackageBlueprint-P3 & with & 1.0 & 1.0 & 0.0 & 0.0 & 0.0 & 0.0 & 0.0 & 0.0 \\
\hline
Moose-TestResources-PackageBlueprint-P4 & without & 1.0 & 1.0 & 0.0 & 0.0 & 0.0 & 0.0 & 0.0 & 0.0 \\
Moose-TestResources-PackageBlueprint-P4 & with & 1.0 & 1.0 & 0.0 & 0.0 & 0.0 & 0.0 & 0.0 & 0.0 \\
\hline
Moose-TestResources-Reference-Core & without & 0.463 & 0.33 & 0.35 & 0.41 & 0.45 & 0.53 & 0.61 & 0.68 \\
Moose-TestResources-Reference-Core & with & 0.464 & 0.37 & 0.37 & 0.4 & 0.45 & 0.51 & 0.59 & 0.65 \\
\hline
Moose-TestResources-Reference-External & without & 0 & 0.0 & 0.0 & 0.0 & 0.0 & 0.0 & 0.0 & 0.0 \\
Moose-TestResources-Reference-External & with & 0 & 0.0 & 0.0 & 0.0 & 0.0 & 0.0 & 0.0 & 0.0 \\
\hline
Moose-TestResources-Reference-PackageOne & without & 0 & 0.0 & 0.0 & 0.0 & 0.0 & 0.0 & 0.0 & 0.0 \\
Moose-TestResources-Reference-PackageOne & with & 0 & 0.0 & 0.0 & 0.0 & 0.0 & 0.0 & 0.0 & 0.0 \\
\hline
Moose-TestResources-Reference-PackageTwo & without & 0 & 0.0 & 0.0 & 0.0 & 0.0 & 0.0 & 0.0 & 0.0 \\
Moose-TestResources-Reference-PackageTwo & with & 0 & 0.0 & 0.0 & 0.0 & 0.0 & 0.0 & 0.0 & 0.0 \\
\hline
Moose-WelcomeBrowser & without & 0.410 & 0.0 & 0.22 & 0.33 & 0.36 & 0.63 & 0.65 & 0.69 \\
Moose-WelcomeBrowser & with & 0.284 & 0.0 & 0.1 & 0.17 & 0.18 & 0.49 & 0.51 & 0.53 \\
\hline
MooseIDE-Analysis & without & 0.193 & 0.01 & 0.03 & 0.16 & 0.21 & 0.25 & 0.3 & 0.4 \\
MooseIDE-Analysis & with & 0.380 & 0.08 & 0.31 & 0.36 & 0.41 & 0.47 & 0.48 & 0.56 \\
\hline
MooseIDE-AttributedText & without & 0.288 & 0.01 & 0.1 & 0.28 & 0.33 & 0.32 & 0.59 & 0.72 \\
MooseIDE-AttributedText & with & 0.348 & 0.06 & 0.17 & 0.33 & 0.38 & 0.5 & 0.59 & 0.72 \\
\hline
MooseIDE-ButterflyMap & without & 0.523 & 0.2 & 0.45 & 0.54 & 0.63 & 0.61 & 0.62 & 0.68 \\
MooseIDE-ButterflyMap & with & 0.435 & 0.25 & 0.32 & 0.44 & 0.53 & 0.49 & 0.5 & 0.55 \\
\hline
MooseIDE-ButterflyMap-Tests & without & 0.345 & 0.22 & 0.27 & 0.27 & 0.45 & 0.49 & 0.45 & 0.48 \\
MooseIDE-ButterflyMap-Tests & with & 0.336 & 0.22 & 0.26 & 0.26 & 0.43 & 0.47 & 0.44 & 0.47 \\
\hline
MooseIDE-ClassBlueprint & without & 0.219 & 0.03 & 0.06 & 0.2 & 0.23 & 0.83 & 0.83 & 1.0 \\
MooseIDE-ClassBlueprint & with & 0.244 & 0.09 & 0.12 & 0.2 & 0.23 & 0.83 & 0.83 & 1.0 \\
\hline
MooseIDE-CoUsageMap & without & 0.329 & 0.02 & 0.11 & 0.29 & 0.44 & 0.49 & 0.53 & 0.58 \\
MooseIDE-CoUsageMap & with & 0.328 & 0.02 & 0.1 & 0.27 & 0.44 & 0.49 & 0.53 & 0.59 \\
\hline
MooseIDE-CoUsageMap-Tests & without & 0.200 & 0.03 & 0.08 & 0.12 & 0.26 & 0.26 & 0.27 & 0.43 \\
MooseIDE-CoUsageMap-Tests & with & 0.149 & 0.03 & 0.05 & 0.07 & 0.19 & 0.19 & 0.19 & 0.35 \\
\hline
MooseIDE-Core & without & 0.264 & 0.03 & 0.09 & 0.26 & 0.29 & 0.36 & 0.4 & 0.45 \\
MooseIDE-Core & with & 0.283 & 0.05 & 0.15 & 0.28 & 0.31 & 0.37 & 0.41 & 0.45 \\
\hline
MooseIDE-Core-Reporter & without & 0.254 & 0.01 & 0.01 & 0.26 & 0.36 & 0.37 & 0.39 & 0.41 \\
MooseIDE-Core-Reporter & with & 0.287 & 0.03 & 0.12 & 0.28 & 0.38 & 0.39 & 0.41 & 0.43 \\
\hline
MooseIDE-CriticBrowser & without & 0.249 & 0.04 & 0.11 & 0.22 & 0.27 & 0.3 & 0.31 & 0.5 \\
MooseIDE-CriticBrowser & with & 0.267 & 0.06 & 0.12 & 0.22 & 0.3 & 0.32 & 0.33 & 0.52 \\
\hline
MooseIDE-CriticBrowser-Tests & without & 0.165 & 0.0 & 0.07 & 0.14 & 0.16 & 0.16 & 0.18 & 0.45 \\
MooseIDE-CriticBrowser-Tests & with & 0.156 & 0.0 & 0.07 & 0.13 & 0.15 & 0.15 & 0.16 & 0.43 \\
\hline
MooseIDE-Dependency & without & 0.481 & 0.12 & 0.34 & 0.52 & 0.6 & 0.6 & 0.64 & 0.66 \\
MooseIDE-Dependency & with & 0.384 & 0.12 & 0.23 & 0.42 & 0.49 & 0.47 & 0.5 & 0.53 \\
\hline
MooseIDE-Duplication & without & 0.420 & 0.01 & 0.39 & 0.43 & 0.49 & 0.52 & 0.53 & 0.6 \\
MooseIDE-Duplication & with & 0.161 & 0.02 & 0.05 & 0.16 & 0.2 & 0.22 & 0.22 & 0.29 \\
\hline
MooseIDE-Export & without & 0.343 & 0.06 & 0.26 & 0.35 & 0.4 & 0.43 & 0.44 & 0.46 \\
MooseIDE-Export & with & 0.288 & 0.1 & 0.19 & 0.28 & 0.33 & 0.36 & 0.37 & 0.39 \\
\hline
MooseIDE-Famix & without & 0.277 & 0.02 & 0.1 & 0.2 & 0.32 & 0.39 & 0.43 & 0.5 \\
MooseIDE-Famix & with & 0.309 & 0.07 & 0.21 & 0.26 & 0.34 & 0.39 & 0.43 & 0.49 \\
\hline
MooseIDE-LayerVisualization & without & 0.420 & 0.15 & 0.27 & 0.4 & 0.54 & 0.54 & 0.55 & 0.64 \\
MooseIDE-LayerVisualization & with & 0.420 & 0.16 & 0.26 & 0.4 & 0.54 & 0.54 & 0.55 & 0.64 \\
\hline
MooseIDE-Meta & without & 0.231 & 0.02 & 0.07 & 0.24 & 0.29 & 0.3 & 0.32 & 0.41 \\
MooseIDE-Meta & with & 0.246 & 0.04 & 0.09 & 0.25 & 0.3 & 0.31 & 0.34 & 0.42 \\
\hline
MooseIDE-NewTools & without & 0.241 & 0.02 & 0.11 & 0.22 & 0.31 & 0.32 & 0.34 & 0.38 \\
MooseIDE-NewTools & with & 0.254 & 0.05 & 0.15 & 0.23 & 0.31 & 0.33 & 0.34 & 0.38 \\
\hline
MooseIDE-NewTools-Tests & without & 0.203 & 0.0 & 0.06 & 0.17 & 0.17 & 0.24 & 0.26 & 0.53 \\
MooseIDE-NewTools-Tests & with & 0.123 & 0.0 & 0.0 & 0.04 & 0.04 & 0.12 & 0.2 & 0.47 \\
\hline
MooseIDE-QueriesBrowser & without & 0.329 & 0.06 & 0.16 & 0.36 & 0.38 & 0.42 & 0.45 & 0.49 \\
MooseIDE-QueriesBrowser & with & 0.349 & 0.08 & 0.19 & 0.37 & 0.4 & 0.45 & 0.47 & 0.51 \\
\hline
MooseIDE-QueriesBrowser-Tests & without & 0.463 & 0.02 & 0.34 & 0.52 & 0.56 & 0.57 & 0.59 & 0.64 \\
MooseIDE-QueriesBrowser-Tests & with & 0.449 & 0.06 & 0.33 & 0.51 & 0.54 & 0.55 & 0.56 & 0.61 \\
\hline
MooseIDE-QueriesDashboard & without & 0.694 & 0.53 & 0.56 & 0.66 & 0.75 & 0.78 & 0.79 & 0.83 \\
MooseIDE-QueriesDashboard & with & 0.694 & 0.53 & 0.56 & 0.66 & 0.74 & 0.78 & 0.78 & 0.82 \\
\hline
MooseIDE-Spotter & without & 0.310 & 0.0 & 0.03 & 0.09 & 0.38 & 0.41 & 0.56 & 0.71 \\
MooseIDE-Spotter & with & 0.3505 & 0.13 & 0.13 & 0.15 & 0.38 & 0.41 & 0.56 & 0.71 \\
\hline
MooseIDE-Spotter-Tests & without & 0.105 & 0.0 & 0.0 & 0.0 & 0.05 & 0.08 & 0.16 & 0.45 \\
MooseIDE-Spotter-Tests & with & 0.146 & 0.07 & 0.07 & 0.07 & 0.07 & 0.1 & 0.19 & 0.45 \\
\hline
MooseIDE-Tagging & without & 0.214 & 0.0 & 0.02 & 0.26 & 0.28 & 0.3 & 0.32 & 0.33 \\
MooseIDE-Tagging & with & 0.227 & 0.01 & 0.04 & 0.26 & 0.29 & 0.32 & 0.34 & 0.34 \\
\hline
MooseIDE-Tagging-Tests & without & 0.231 & 0.0 & 0.12 & 0.17 & 0.18 & 0.37 & 0.41 & 0.44 \\
MooseIDE-Tagging-Tests & with & 0.206 & 0.0 & 0.1 & 0.14 & 0.15 & 0.33 & 0.38 & 0.42 \\
\hline
MooseIDE-Tests & without & 0.205 & 0.04 & 0.15 & 0.18 & 0.21 & 0.25 & 0.26 & 0.39 \\
MooseIDE-Tests & with & 0.185 & 0.05 & 0.13 & 0.16 & 0.19 & 0.23 & 0.24 & 0.36 \\
\hline
MooseIDE-Visualization & without & 0.539 & 0.1 & 0.42 & 0.56 & 0.61 & 0.67 & 0.73 & 0.79 \\
MooseIDE-Visualization & with & 0.328 & 0.09 & 0.17 & 0.34 & 0.39 & 0.41 & 0.46 & 0.52 \\

\end{longtable}

\end{document}